\documentstyle[preprint,version2,aps]{revtex}  % use this one
\begin{document}
\draft
\preprint{OCIP/C 95-3}
\preprint{UQAM-PHE-95-08}
\preprint{April 1995}
\begin{title}
W-PAIR PRODUCTION IN THE PROCESS $e^+e^- \to \ell \nu q\bar{q}'$ \\
AND MEASUREMENT OF THE $WW\gamma$ and $WWZ$ COUPLINGS
\end{title}
\author{Mikul\'{a}\v{s} Gintner and Stephen Godfrey\footnote{e-mail:
godfrey@physics.carleton.ca}}
\begin{instit}
Ottawa-Carleton Institute for Physics \\
Department of Physics, Carleton University, Ottawa CANADA, K1S 5B6
\end{instit}
\moreauthors{Gilles Couture\footnote{e-mail: couture@osiris.phy.uqam.ca}}
\begin{instit}
D\'epartement de  Physique, Universit\'e du Qu\'ebec \`a Montr\'eal \\
C.P. 8888, Succ. Centre-Ville, Montr\'eal, Qu\'ebec, Canada, H3C 3P8
\end{instit}
\begin{abstract}
We performed a detailed analysis of the process
$e^+e^-\to \ell \nu q\bar{q}'$ where we included
all tree level Feynman diagrams that
contribute to this final state. We studied the sensitivity of this
process to anomalous trilinear gauge boson couplings of the $WW\gamma$
and $WWZ$ vertices using two popular parametrizations.  We used a
maximum likelihood analysis of a five dimensional differential
cross-section based on the $W$ and $W$ decay product angular
distributions.
We concentrated on
LEP-200 energies, taking $\sqrt{s}=175$ GeV, and energies appropriate to the
proposed Next Linear Collider (NLC), a high energy $e^+e^-$ collider
with center of mass energies $\sqrt{s}=500$
and 1~TeV. At 175 GeV, $g_1^Z$ can be measured to about $\pm 0.2$,
$\kappa_Z$ to $\pm 0.2$ and $\kappa_\gamma$ to $\pm 0.3$,
$\lambda_Z$ to $\pm 0.2$ and $\lambda_\gamma$ to $\pm 0.3$.
at 95\% C.L. assuming 500~pb$^{-1}$ integrated
luminosity.  Although these will be improvements of existing
measurements they are not sufficiently precise to test the standard
model at the loop level and are unlikely to see deviations from SM
expectations.  At 500~GeV with 50~fb$^{-1}$ integrated
luminosity, $g_1^Z$ can be measured to about $\pm 0.01$,
$\kappa_Z$ and $\kappa_\gamma$ to $\pm 0.005$
and $\lambda_Z$ and $\lambda_\gamma$ to $\pm 0.003$ at 95\% C.L.
while at 1 TeV with 200~fb$^{-1}$ integrated
luminosity, $\kappa_V$ and $\lambda_V$
can be measured to about $\pm 0.005$ and $\pm
10^{-3}$ respectively.
The 500~GeV measurements will be at the level of loop
contributions to the couplings and may show hints of new physics while
the 1~TeV  should be sensitive to new physics at the loop level.
\end{abstract}

\narrowtext
\section{INTRODUCTION}
\label{sec:intro}

$e^+e^-$ colliders have made important contributions to our
understanding of the
electroweak interactions\cite{lep} and it is expected that
this tradition will continue with the advent of higher luminosity
and higher energy machines.
In the near future
the CERN LEP-200 $e^+e^-$ collider\cite{hagiwara87,lep200,kane89}
will begin operation and beyond that there is a growing effort
directed towards the design and construction of future high energy $e^+e^-$
linear colliders with $\sqrt{s}\geq 500$~GeV
which we will refer to generically as the Next
Linear Collider (NLC)\cite{nlcphysics,miyamoto,nlc,JLC,NLC2,CLIC}.
One of the primary physics goals of LEP-200
and an important goal of the NLC is to
make precision measurements of  $W$ boson properties including
$M_W$, $\Gamma_W$, and $W$-boson interactions with
fermions and the photon and $Z^0$.

The latter measurements, that of
the trilinear gauge boson vertices (TGV's) provides a
stringent test of the gauge structure of the standard
model\cite{aihara95,tgvreviews}.
The current measurement of these couplings are rather weak.
Using a popular parametrization of the CP conserving gauge
boson couplings, indirect measurements of TGV's via radiative corrections to
precision electroweak measurements\cite{burgess94,dawson95,hagiwara93}
give the following limits\cite{burgess94}:
$\delta g_Z^1 =-0.033\pm 0.031$, $\delta \kappa_\gamma =0.056\pm
0.056$, $\delta \kappa_Z =-0.0019\pm 0.044$,
$\lambda_\gamma=-0.036\pm 0.034 $, and $\lambda_Z =0.049\pm 0.045$.
However, there are ambiguities in these calculations associated with
running the couplings down from the scale of new physics to low
energy so that these
limits are not particularly rigorous and it is necessary
to use direct measurements for more reliable bounds.
The CDF and D0 collaborations at the Tevatron
$p\bar{p}$ collider at Fermilab, using the processes
$p\bar{p} \to W \gamma , \; WW, \; WZ$,
have obtained the direct 95\% C.L. limits of
$-1.6 <\delta\kappa_\gamma < 1.8$, $-0.6 < \lambda_\gamma < 0.6$,
$-8.6<\delta\kappa_Z < 9.0$, and $-1.7 < \lambda_Z < 1.7$
\cite{tevatron}.
These measurements are quite weak but
it is expected that they will improve as the
luminosity of the Tevatron increases.  In the longer term
measurements at the Large Hadron Collider at CERN will
improve these limits considerably\cite{lhc}.

It is expected that measurements at high energy $e^+e^-$ colliders will
surpass those at the hadron colliders.  As a result,
many processes have been studied to determine
their usefulness for measuring the TGV's;
$e\gamma\to \nu W$ \cite{yehudai,godfrey92,couture,egamma},
$e^-e^- \to e^- W^-\nu_e$ \cite{cuypers93},
$\gamma\gamma \to W^+ W^-$\cite{yehudai,choi91},
$e^+ e^- \to Z \nu \bar{\nu}$ \cite{couture92,ambrosanio92,hagiwara91},
$e^+e^-\to \gamma \nu\bar{\nu}$\cite{miyamoto,couture94,eptog},
$e^+e^-\to W^+W^-$
\cite{hagiwara87,aihara95,barklow92,papa95,sekulin,eptoww,pankov95}
and more detailed studies of various four fermion final states in
the process $e^+e^- \to W^+ W^- \to f\bar{f} f'\bar{f}'$
\cite{beenakker94,couture92,kalyniak93,gintner95}.

Probably the most useful of the $e^+e^-\to W^+W^-$ channels
for these studies is $e^+e^- \to \ell
\nu q \bar{q}'$.
With only one unobserved neutrino this channel has several advantages:
it can be fully reconstructed
using the constraint of the initial beam energies,
the  $W^+$ and $W^-$ can be discriminated using lepton charge identification,
it does not have the QCD backgrounds that plague the fully hadronic
decay modes, and it offers much higher statistics than the fully
leptonic modes.
As a result of the importance of this channel there have been
numerous studies of this process.  In particular, there is a growing
list of analysis of $e^+e^- \to W^+W^-$,
$e^+e^-$ to four fermion final state processes
\cite{aihara95,couture92,kalyniak93,papa95,kurihara94,fourfermions},
single $W$ production\cite{hagiwara91,singlew}, electroweak
radiative corrections to these reactions including the important
contribution from initial state radiation
\cite{emrc,berends94b,beenakker91,fleischer94,oldenborgh94,bardin94},
and the sensitivity of these
processes to anomalous $WW\gamma$ and $WWZ^0$ gauge boson couplings (TGV's).

In this paper we examine in detail the four fermion final state
$e^+e^- \to \ell \nu_\ell q \bar{q}'$ where $\ell$ is either $e^\pm$
or $\mu^\pm$ and $q\bar{q}'$ can be either $(ud)$ or $(cs)$.
We study this process for the centre of mass energies
$\sqrt{s}=175$~GeV appropriate to LEP200, and $\sqrt{s}=500$ and
1000~GeV appropriate to the NLC.
To obtain results we included all tree level diagrams to the four
fermion final states using helicity amplitude techniques.  For the
$\mu^\pm \nu_\mu q \bar{q}'$ final state 10 diagrams contribute and
for the $e^\pm \nu_e q\bar{q}'$ final state 20 diagrams contribute.
Our primary interest is to study the sensitivity of these processes
to anomalous gauge boson couplings.  To do so we examined numerous
distributions.  For the purpose of comparing theory to experiment we
also examined the question of whether  the approximation
of only including the resonant diagrams is adequate or whether the
full four-fermion final state calculation is needed.  Using helicity
amplitudes we are able to study the usefulness of initial state
polarization in extracting the TGV's.  In the $e^\pm \nu q\bar{q}'$
final state single $W$ production can also be studied\cite{singlew}
where the $WW\gamma$ vertex can be isolated from the $WWZ$ vertex by
imposing an appropriate cut on the outgoing electron.  In this case,
when only hadronic jets are observed and not the outgoing lepton,
there are ambiguities in identifying the charge of the $W$ besides
problems with hadronic backgrounds which we do not deal with here.
A detailed analysis of single $W$ production will be
presented elsewhere \cite{gintner95b}.

In the next section we discuss effective Lagrangians and the
various parametrizations used to describe
gauge boson self interactions which have appeared in the
literature.  In section III we describe our calculation.  Section IV
comprises the bulk
of the paper which is used to present and discuss our results.  We
summarize our conclusions in section V.

\section{PARAMETRIZATION OF THE TRIPLE GAUGE BOSON COUPLINGS}

The formalism of effective Lagrangians provides a well-defined
framework for investigating the physics of anomalous couplings and
electroweak symmetry breaking \cite{aihara95,boudjema,bagger93}.
In this approach
an infinite set of non-renormalizable operators, consistent with the
unbroken symmetries and whose coefficients
parametrize the low-energy effects of electroweak symmetry breaking
or new physics are organized in an energy expansion.
At low energy only a
finite number of terms contribute to a given process.  At
higher energies more and more terms become important until the
whole process breaks down at the scale of new physics. One focuses
on the leading operators in the expansion.

There are three main parametrizations of gauge boson couplings that
appear in the
literature.  The characteristic distinguishing the
approaches is the degree to which constraints are imposed in terms
of the symmetry and particle content of the low energy theory.  We
summarize the most commonly used parametrizations
below\cite{aihara95,boudjema}.

\subsection{General Form Factor Approach}

The first approach is to describe
the $WWV$ verticies using the most general
parametrization possible that respects Lorentz invariance,
electromagnetic gauge invariance and $CP$ invariance
\cite{hagiwara87,gaemers79,miscvertex}. This approach
has become the standard
parametrization used in phenomenology making the comparison
of the sensitivity of different measurements to the TGV's straightforward.
We do not
consider CP violating operators in this paper as they  are tightly
constrained by measurement of the neutron
electric dipole moment which constrains the two CP violating parameters to
$|\tilde{\kappa}|, |\tilde{\lambda}|< {\cal O} (10^{-4})$ \cite{cp}.
With these constraints the $WW\gamma$ and $WWZ$ vertices have five
free independent parameters,
$g_1^Z$, $\kappa_\gamma$, $\kappa_Z$, $\lambda_\gamma$ and $\lambda_Z$ and
is given by \cite{hagiwara87,gaemers79}:
\begin{equation}
{\cal L}_{WWV} =  - ig_V \left\{ { g_1^V (W^+_{\mu\nu}W^{-\mu} -
W^{+\mu}  W_{\mu\nu} ) V^\nu
+ \kappa_V W^+_\mu W^-_\nu V^{\mu\nu}
+ {{\lambda_V}\over{M_W^2}} W^+_{\lambda\mu}W^{-\mu}_\nu V^{\nu\lambda}
}\right\}
\end{equation}
where the subscript $V$ denotes either a photon or a $Z^0$,
$V^\mu$ and $W^\mu$ represents  the photon or $Z^0$ and $W^-$
fields respectively,
$W_{\mu\nu}=\partial_\mu W_\nu-\partial_\nu W_\mu$ and
$V_{\mu\nu}=\partial_\mu V_\nu-\partial_\nu V_\mu$
and $M_W$ is the $W$ boson mass.  ($g_1^\gamma$ is constrained by
electromagnetic gauge invariance to be equal to 1.)
The first two terms correspond to dimension 4 operators and the
third term corresponds to a dimension 6 operator.  The mass in the
denominator of the dimension 6 term would correspond to the scale of
new physics, typically of order 1~TeV.  However, it has become the
convention to use $M_W$ so that the $W$ magnetic
dipole and electric quadrupole can be written in a form similar to
that of the muon.  Nevertheless, one expects the dimension 6
operator to be suppressed with respect to the dimension 4 operators by
a factor of $M_W^2/(\Lambda=1\;\hbox{TeV})^2 \simeq 10^{-2}$.
Higher dimension operators correspond to momentum
dependence\cite{baur88} in the form
factors which are not so important in the process we are considering so are
not included.
At tree level the standard model (SM) requires $g_1^Z=\kappa_V=1$ and
$\lambda_V=0$.
Typically, radiative corrections from heavy particles will change
$\kappa_V$ by about $\sim 10^{-2}$ and $\lambda_V$ by about $\sim
10^{-3}$ \cite{smloops}.  In particular, the contributions from a
200~GeV top quark and a 150~GeV Higgs boson to $\kappa_V$ and
$\lambda_V$ are of order $10^{-3}$.

The nearness of the $\rho$ parameter to 1 implies an $SU(2)$
invariance of the weak interaction.  Imposing this from the outset
implies a relationship between the parameters reducing the number of
parameters from 5 to 3 \cite{kuroda87}.  $SU(2)$ invariance
is the basis of the other two parametrizations we mention.  The
difference between the two is that in the first, the Higgs boson is
heavy so that Goldstone bosons are nonlinearly realized while in
the second, the Higgs bosons are light, leading to linearly
realized Goldstone bosons.

\subsection{Non-Linearly Realized Higgs Sector}

The second commonly used parametrization  is
the Chiral Lagrangian approach\cite{bagger93,nonlinear}.
A custodial $SU(2)$ is assumed which is supported to high accuracy by the
nearness of the $\rho$ parameter to 1. This approach assumes that the
theory has no light Higgs particles and the electroweak gauge bosons
interact strongly with each other above approximately 1~TeV.  This
can be described by a non-linear realization of the $SU(2)\times
U(1)$ symmetry in a chiral Lagrangian formalism leading to the
effective Lagrangian:
\begin{equation}
L= -i g {{L_{9L}}\over{16\pi^2}} Tr [W^{\mu\nu} D_\mu \Sigma
D_\nu \Sigma^\dagger  ]
-i g' {{L_{9R}}\over{16\pi^2}} Tr [B^{\mu\nu} D_\mu \Sigma^\dagger
D_\nu \Sigma ]
+ g g'{{L_{10}}\over{16\pi^2}} Tr [ \Sigma B^{\mu\nu} \Sigma^\dagger
W_{\mu\nu}]
\end{equation}
where $W_{\mu\nu}$ and $B_{\mu\nu}$ are the $SU(2)$ and $U(1)$ field
strength tensors given in terms of $W_\mu\equiv W_\mu^i \tau_i$ by
\begin{eqnarray}
W_{\mu\nu} & = & {1\over 2} (\partial _\mu W_\nu -\partial_\nu W_\mu
+{i\over 2} g[W_\mu , W_\nu ]) \nonumber\\
B_{\mu\nu} & = & {1\over 2} (\partial _\mu B_\nu -\partial_\nu B_\mu )\tau_3 ,
\end{eqnarray}
$\Sigma=\exp(iw^i\tau^i/v)$,  $v=246$~GeV, $w^i$ are the would-be
Goldstone bosons that give the $W$ and $Z$ their masses via the
Higgs mechanism, and the $SU(2)_L \times U(1)_Y$ covariant
derivative is given by
$D_\mu \Sigma = \partial_\mu \Sigma + {1\over 2} i g W_\mu^i \tau^i \Sigma
-{1\over 2} i g' B_\mu \Sigma \tau^3 $.  The Feynman rules are found
by going to the unitary gauge where $\Sigma=1$.  Note that often in
the literature the coefficient $1/16\pi^2$ is replaced with $v^2/\Lambda^2$.
$L_{10}$ contributes to the gauge boson self energies where it is tightly
constrained to $-1.1 \leq L_{10} \leq 1.5$ \cite{dawson95}
so we will not consider it
further.  New physics contributions are expected to result in values
of $L_{9L,9R}$ of order 1\cite{bagger93}.

\subsection{Linearly Realized Higgs Sector}

In the linear realization scenario\cite{linear} the Higgs doublet field $\Phi$
is included in the low energy particle content.
This approach assumes that any deviations
from the Standard Model due to new physics manifests themselves in
$SU(3)\times SU(2) \times U(1)$ gauge invariant singlet operators.
There are 7 relevant operators of which four
are stringently constrained by the high precision
low energy and $Z$ boson data \cite{hagiwara93}. The remaining three
can give rise to non-standard couplings
\begin{eqnarray}
{\cal L} & = & ig' {\varepsilon_B \over \Lambda^2} (D_\mu
\Phi)^\dagger B^{\mu\nu} (D_\nu \Phi ) + ig {\varepsilon_W\over \Lambda^2}
(D_\mu \Phi)^\dagger W^{\mu\nu} (D_\nu \Phi ) \\
& & \qquad\quad + {2i\over 3} {L_\lambda \over \Lambda^2} g^3 Tr [W_{\mu\nu}
W^{\nu\rho} W^\mu_\rho ]
\end{eqnarray}
It seems most likely that
anomalous couplings in the light Higgs linear scenario would best
be studied by measuring the properties and couplings of the Higgs
boson directly.  In any case the parameters from this approach can
be rewritten in terms of the parameters of the first two Lagrangians
discussed.

The parameters from the three Lagrangians can be mapped onto each other:
\begin{displaymath}
\begin{array}{lll}
g_1^Z & = 1 +{{e^2}\over {32\pi^2 s^2_w c^2_w}} (L_{9L} + {{2s^2
L_{10}}\over{(c^2_w-s^2_w)}}) & =1 + {e^2\over s_w^2} {v^2\over
4 \Lambda^2} ({\varepsilon_W \over c_w^2}) \\
\kappa_z & = 1 + {{e^2}\over {32\pi^2 s^2_w c^2_w}}
(L_{9L}c^2_w - L_{9R} s^2_w)
 +{{4s^2_w c^2_w}\over {(c^2_w-s^2_w)}} L_{10}
& = 1+ {e^2 \over s_w^2} {v^2\over {4\Lambda^2}} (\varepsilon_W
-{s_w^2\over c_w^2} \varepsilon_B) \\
\kappa_\gamma & = 1 + {1\over 32\pi^2} {e^2\over s^2_w} (L_{9L} +
L_{9R} - 2L_{10}) & = 1 +{e^2\over s_w^2} {v^2\over 4\Lambda^2}
(\varepsilon_W + \varepsilon_B) \\
\lambda_\gamma & = \lambda_z & = ({e^2\over s_w^2}) L_\lambda
{M_W^2\over \Lambda^2}
\end{array}
\end{displaymath}
Dropping the $L_{10}$ term, the linear and non-linear realizations
are obtained from each other by identifying $L_{9L}=2\varepsilon_W$
and $L_{9R}=2\varepsilon_B$.
In the non-linear realization, the counterpart of $L_\lambda$ is
higher dimension.

\section{CALCULATIONS AND RESULTS}

To study the process $e^+e^-\to \ell^\pm \nu q \bar{q}'$
we included all tree level diagrams to the four
fermion final states.  There are
10 diagrams contributing to the $e^+e^- \to \mu^\pm
\nu_\mu q \bar{q}'$ final state which are shown in Fig. 1.
The gauge boson coupling we are studying is present in diagram (1a).
This, along with diagram (1b) are the diagrams responsible for real
$W$ production.
For the $e^\pm \nu_e q\bar{q}'$ final state the 10 diagrams shown in
Fig. 2 must also be included with those of Fig. 1 for a total
of 20 diagrams.
Diagram (2a) includes a TGV.  The diagrams with
t-channel photon exchange make large contributions to single $W$
production due to the pole in the photon propagator which can be
used to isolate the $WW\gamma$ vertex from the $WWZ$ vertex
\cite{singlew,gintner95b}.

We include final width effects by using
vector boson propagators of the form $(s-M_V^2 +
i\Gamma_V M_V)^{-1}$ which yields a gauge invariant result.
Strictly speaking we should have included a momentum dependent
vector boson width but this leads to problems with gauge
invariance\cite{papa95,kurihara94,baur95,zeppenfeld92}.
Although a number of solutions to this problem have been discussed
\cite{papa95,kurihara94,baur95} the difference
between our treatment and more rigorous ones have a totally negligible
effect on the TGV sensitivities we obtain from our analysis.  A
more rigorous treatment must of course be included in Monte Carlo
simulations that will be used to analyze real experimental data.
We will find that the non-resonant diagrams make non-negligible
contributions to cross sections and are dependent on the kinematic
cuts used in the analysis. These contributions are at least as
important as electroweak radiative corrections.

To evaluate the cross-sections and different distributions, we used the CALKUL
helicity amplitude technique \cite{calkul} to obtain expressions for the
matrix elements and
performed the phase space integration using  Monte Carlo
techniques \cite{monte}.
The expressions for the helicity amplitudes are lengthy and unilluminating so
we do not include them here.
%The interested reader can obtain them directly
%from the authors.
To obtain numerical results we used the values $\alpha=1/128$,
$\sin^2\theta=0.23$, $M_Z=91.187$ GeV, $\Gamma_Z=2.49$ GeV,
$M_W=80.22$ GeV, and
$\Gamma_W=2.08$ GeV.  In our results we included two generations of
quarks and took them to be massless.
In order to take into account finite detector acceptance we require
that the lepton and quarks are at least
10 degrees away from the beam and have at least 10~GeV energy unless
otherwise noted.
%We also
%require that there be a missing transverse momentum ($\not{p}_T$) of
%at least 10~GeV.  We will discuss single $W$ production separately.

In principle we should include QED radiative corrections from soft photon
emission and the backgrounds due to a photon that is lost
down the beam pipe \cite{emrc,berends94b,fleischer94}.
These backgrounds are well
understood and detector dependent.  We assume the approach taken at
LEP, that these effects can best be taken into account by the
experimental collaborations.  In any case, although initial state radiation
must be taken into account their inclusion does not
substantially effect the bounds we obtain and therefore our conclusions.

In figure 3 and Table I we show the cross sections for the processes $e^+e^-
\to \ell^\pm \nu_\ell q \bar{q}'$ for different applications of cuts
on the invariant mass of the lepton-neutrino pair and the
$q\bar{q}'$ pair; $|M_{(l\nu),(q\bar{q}')} - M_W|<5$~GeV where
$M_{(\ell\nu),(q\bar{q}')}$ is the invariant mass of the $\ell\nu$
and $q\bar{q}'$ pair respectively.  Imposing the cut on one fermion
pair gives the single $W$ cross section and imposing the cut on both
fermion pairs gives the $W$-pair production cross section.
In both cases the single $W$ and $W$ pair thresholds are clearly seen.
Although the single $W$ production cross section is nonzero below
the $W$-pair production threshold, it is still
too small to obtain adequate statistics
to perform studies of $W$ boson properties.
In general the electron mode has a larger cross-section than the
muon mode.  The difference is small at 175~GeV but becomes
increasingly larger at higher energy as the t-channel photon
exchange  becomes increasingly important, reaching a factor of 5 at 1~TeV.
For the muon mode the invariant mass cuts reduce the
cross-section by 10\% to 20\% depending on $\sqrt{s}$
irrespective of whether the cut is on $M_{\ell\nu}$ or $M_{q\bar{q}}$.
The relatively small effect of these cuts
verifies the dominance of the resonant diagrams on the total cross
section.

For the electron mode
the results are similar when the $|M_{e\nu} - M_W|<5$~GeV cut is
imposed which constrains the $e\nu$ pair to be on the $W$ mass-shell.
However when $|M_{q\bar{q}'} -
M_W|<5$~GeV (ie. $W\to q\bar{q}'$) the cross section is
significantly larger than the previous case due to
the enhancement arising from the non-resonant
t-channel photon exchange diagrams of fig. 2.  With appropriate
kinematic cuts this can be used to study single-$W$ production
\cite{gintner95b}.

Despite the relative smallness of the
off-resonance contributions to the muon mode they still contribute
up to 30\% of the cross section at 1~TeV.  Clearly, they must be
properly included when making high precision tests of standard model
processes.  For the electron mode they are even more important and
are interesting in the context of single $W$ production.

\subsection{Distributions}

The above points can be amplified by examining kinematic
distributions.  In addition, since our goal is to extract
measurements of the TGV's, we must explore which distributions
are most sensitive to anomalous couplings.  For descriptive purposes
we will show various distributions for $\sqrt{s}=500$~GeV.

We begin by showing in Fig. 4
the invariant mass distributions for the $e\nu$ ($M_{e\nu}$)
and $\mu \nu$ ($M_{\mu\nu}$) pairs
for left and right handed initial electron polarization.
As the unpolarized
cross sections are dominated by the left handed electrons they are
quite similar to them, so we do not
include them separately.  In addition, the $q\bar{q}$ invariant mass
distributions for left handed initial electrons are similar to
the $M_{\mu\nu}$ distributions.  Since there are differences
for the right handed initial electron distributions, these
distributions are also included.  The differences in these cross
sections reflects the differences and relative importance
in the Feynman diagrams that contribute to a process.
Although the cross
sections and the sensitivities to the TGV's are dominated by the
production of real $W$'s one can see that off-resonance
production of the $\ell\nu q\bar{q}'$ final state can be quite
sensitive to anomalous couplings.  We will explore this in
a later section.
The effects are especially pronounced for the $e\nu q\bar{q}'$
final state where there is the possibility of single $W$
production which is discussed elsewhere \cite{singlew,gintner95b}.

We examined numerous distributions with the purpose of
finding the distributions and isolating the regions of phase space
most sensitive to anomalous couplings;
\begin{equation}
{{d\sigma}\over {dp_{T_W}}}, \quad {{d\sigma}\over {dp_{T_\mu}}},
\quad {{d\sigma}\over {dp_{E_\mu}}}, \quad
{{d\sigma}\over {d\cos\theta_{eW}}}, \quad
{{d\sigma}\over {d\cos\theta_{e\mu}}}\quad \hbox{etc.}
\end{equation}
There is, of course, overlap among the regions of interest in these
distributions.  To gauge the sensitivity of these distributions to
the TGV's we typically divided them into 4-bins and performed a
$\chi^2$ analysis.  For  $\sqrt{s}=500$~GeV and
integrated luminosity of 50~fb$^{-1}$ we
found, for example, that the $\kappa$'s could be measured to a couple of
percent at 95\% confidence level.  It turns out that this is not
competitive with a more sophisticated analysis of angular
distributions we will describe below.
Generally, this is because the phase space regions
with the highest statistics are least sensitive to anomalous
couplings and tend to overwhelm deviations
while the regions most sensitive to TGV's have poor statistics.

For the purpose of understanding $W$-boson properties the most
interesting distributions are the various angular distributions.  To
understand this better it is useful to first consider the $W$ pair
production cross section without decays to
fermions\cite{peskin88,burke91}.
To leading order the amplitude for $W$ pair-production is given by
three diagrams; via an s-channel photon, an s-channel $Z^0$ and a
t-channel neutrino exchange.
The cross sections at different $\sqrt{s}=200$, 500, and
1000~GeV, for $W_L W_L$,
$W_L W_T$, and $W_T W_T$, different initial state polarizations, and as
a function of the $W$ scattering angle are shown in Fig. 5
\cite{hagiwara87}.
For the initial state $e^-_R e^+_L$
only the first two diagrams contribute which
at high energy is dominated by longitudinal $W$ ($W_L$) production.
Due to the delicate cancellations between the diagrams it is
$W_L$ production which is most sensitive to anomalous couplings.
In contrast, the cross section for the $e^-_L e^+_R$ initial state
produces both transverse and longitudinal $W$ bosons with comparable
rates.
The $e^-_L$ cross section is dominated by a peak in the forward
direction with respect to the incoming $e^-$
associated with the t-channel neutrino exchange which
is made up entirely of transverse $W$ production.  This contribution
is relatively insensitive to new physics. The cross
sections in the backward direction includes sizable longitudinal $W$
production
accounting for about 25\% of the total cross section in
the backward hemisphere.  However, in the backward direction where
the s-channel diagrams contribute substantially, the cross section
for $e^-_R$ is always quite small.  For $e^-_R$ there is a large
change in the magnitude of the cross-section but only a small change
in its shape.

Any disruption of the delicate gauge theory cancellations leads to
large changes to the standard model results.  For
$W_L$ production amplitudes the enhancements can be a factor of
$(s/M_W^2)$. This is shown in Fig. 6 where the angular distribution of the
outgoing $W$ is plotted for several values of anomalous couplings at
$\sqrt{s}=500$~GeV.

Because it is the longitudinal $W$ production which is most
sensitive to anomalous couplings, and because the cross section is
dominated by transverse $W$ production it is crucial to disentangle
the $W_L$ from the $W_T$ {\it background}.    The most convenient
means of doing so makes use of the angular distribution of the $W$
decay products.  Defining $\theta_{\ell}$ and $\theta_q$ as the angle between
the $\ell$ or $q$ and the $W$ momentum measured in the $W$ rest
frame, the angular distribution in $\theta$ peaks about
$\cos\theta=0$ for longitudinally polarized $W$ bosons and at
forward or backward angles for transversely polarized bosons. In
addition the parity violation of the $W$ couplings distinguishes the
two polarization states adding to the effectiveness of the decay as
a polarimeter.   Thus the angular distributions can be used to
extract information about the $W$ boson polarizations.  In Fig. 7 we
show the angular distributions for the outgoing quark with respect
to the $W$ direction ($\theta_q$) for the three bins in the $W$
scattering angle, $\cos\Theta_W <-0.9$, $-0.05<\cos\Theta_W < 0.05$ and
$\cos\Theta_W > 0.9$, (where we take $\Theta_W$ to be the $W^-$ angle
with respect to the incoming $e^-$)
for the process $e^+e^- \to \mu^+\nu_\mu
q\bar{q}'$ at $\sqrt{s}=500$~GeV with the initial electron unpolarized.
Several values of $\kappa_V$ and $\lambda_V$ are included to
demonstrate the sensitivity of the distributions to anomalous
couplings.
The figure shows the dominance of the
transverse $W$ polarization at forward angles and the increasing
importance of the $W$ longitudinal polarization at
$\cos\Theta_W=0$.  Note also the relative lack of sensitivity to
anomalous couplings for the forward, dominantly transverse $W$'s and
how the sensitivity increases as the scattering angle increases and
longitudinal $W$'s contribute a larger fraction of the cross section.

We have shown the $d^2 \sigma /d\cos\Theta_W \; d\cos\theta_q$
distribution as it displays the most dramatic change in the shape of
the distributions.
However, interference
between the transverse and longitudinal $W$'s also depends on the
azimuthal angle so that the azimuthal angular distribution also
shows changes in its shape, albeit smaller.  One finds similar
effects in the angular distributions for the decay $W\to \ell \nu$.

\subsection{Maximum Likelihood fit of 5-dimensional angular
distribution}

The approach which makes the most complete use of information in an event
is the maximum likelihood method.
Based on the observations of the previous section
we perform a maximum likelihood fit
based on the 5 angles\cite{barklow92,sekulin}; $\Theta$, the $W^-$
scattering angle with respect to the initial $e^+$ direction,
$\theta_{qq}$, the polar decay angle of the $q$ in the $W^-$ rest
frame using the $W^-$ direction as the quantization axis,
$\phi_{qq}$, the azimuthal decay angle of the $q$ in the $W^-$ rest
frame, and $\theta_{\ell\nu}$ and $\phi_{\ell\nu}$ are the analogous
angles for the lepton in the $W^+$ rest frame.   The azimuthal
angles are defined as the angle between the normal to the reaction
plane, $n_1=p_e \times p_W$ and the plane defined by the $W$ decay
products, $n_2=p_q \times p_{\bar{q}}$.
The angles are shown in Fig. 8. For the $q\bar{q}$
case there is an ambiguity since we cannot tell which hadronic jet
corresponds to the quark and which to the antiquark.  We therefore
include both possibilities in our analysis.

To implement the maximum likelihood analysis we divided each
of $\Theta$, $\theta_{qq}$, $\phi_{qq}$, $\theta_{\ell\nu}$, and
$\phi_{\ell\nu}$ into four bins so that the entire phase space was
divided into $4^5 = 1024$ bins.  With this many bins some will not
be very populated with events so that it is more appropriate to use
Poisson statistics rather than Gaussian statistics.  This leads
naturally to the maximum likelihood method.
The change in the
log of the Likelihood function
%from the standard model expectation
is given by
\begin{equation}
\delta\ln {\cal L} = \sum [ -r_i +r_i \ln (r_i) +\mu_i -r_i \ln (\mu_i) ]
\end{equation}
where the sum extends over all the bins and $r_i$ and $\mu_i$ are
the predicted number of non-standard model and standard model
events in bin $i$ respectively, given by
\begin{equation}
r_i = L \int_{\Delta\Theta} \int_{\Delta\theta_{qq} }
\int_{\Delta\phi_{qq}} \int_{\Delta_{\ell\nu}}
\int_{\Delta\phi_{\ell\nu}} {{d^5\sigma}\over{d\cos\Theta
d\cos\theta_{qq} d\phi_{qq} d\cos\theta_{\ell\nu} d\phi_{\ell\nu} }}
d\cos\Theta \; d\cos\theta_{\ell\nu} \; d\phi_{\ell\nu} \;
d\cos\theta_{qq} \; d\phi_{qq}
\end{equation}
where $ L$ is the expected integrated luminosity.
The 68\% and 95\% confidence level bounds are given by the values of
anomalous couplings which give a change in $\ln{\cal L}$ of 0.5 and
2.0 respectively.

Ideally, one would perform the analysis on
an event by event basis but to simplify our calculations we used a
five dimensional angular distribution.  To check the sensitivity to
binning we varied the number of dimensions and bins used in our fits.
For this binning approach
we found that the results converged to the tightest bounds
using the five dimensional distribution
and 4 bins per dimension.  In a few test runs for special cases of
kinematic cuts we calculated the likelihood function
on an event by event basis and found that the sensitivities improved
a small amount over the 5 dimensional distribution case described above.

The results we obtained are based solely on the statistical errors based
on the integrated lumininosity we assume for the various cases.
To include the effects of systematic errors using the maximum likelihood
approach requires an unweighted Monte Carlo simulation through a
realistic detector.  Since we did not have the facilities to do this we
attempted a simplified estimate of systematic errors using
$\chi^2$ analysis to make our estimates.
We assumed a systematic error of 5\% of a measurement which we combined
in quadrature with the statistical error.  In general the systematic
errors are negligible compared to the statistical errors.
The only times they
made a measurable difference was for the high luminosity cases of the
500~GeV and 1~TeV NLC, and even there the effect was quite small.  It is
straightforward to see why this is so; with so many bins the number of
events per bin is quite small resulting in a large statistical error.
Thus, it appears that the total errors will be dominated by the
statistical errors but clearly, a full detector Monte Carlo must be
performed to properly understand the situation.

\subsection{Unpolarized Results}

A thorough analysis of gauge boson couplings would allow all five
parameters in the Lagrangian to vary simultaneously to take into
account cancellations (and correlations) among the various contributions.
This approach is  impractical, however, due to the large
amount of computer time that would be required to search the parameter
space.  Instead we show 2-dimensional contours for a selection
of parameter pairs to give a sense of the correlations.
%We believe
%that these contours are reasonably reliable as when
%the other parameters at the edges of our 2-dimensional contours were
%varied, there was little change in the sensitivities.
For the case
of the Chiral Lagrangian where the global SU(2) symmetry imposes
relations between the parameters and where we restrict ourselves to
dimension four operators the parameter space reduces to 2 dimensions.

\subsubsection{$\sqrt{s}=175$~GeV}

%We begin with $\sqrt{s}=200$~GeV.
For the LEP200 collider we study the sensitivity to the gauge boson
couplings using the expected machine parameters of
$\sqrt{s}=175$~GeV and an integrated luminosity of
500~pb$^{-1}$.  These results do not have a cut on $M_{\ell\nu}$ or
$M_{q\bar{q}}$ since for these energies the cuts have
virtually no effect on the
sensitivities except for the electron mode involving the $WW\gamma$
vertex where the effect is still quite small.
The 95\% confidence limits for the $g_Z^1-\kappa_Z$,
$\kappa_\gamma -\kappa_Z$, $\kappa_\gamma - \lambda_\gamma$, and
$\kappa_Z -\lambda_Z$ planes are shown in Fig. 9 and for the
$L_{9L}-L_{9R}$ plane is shown in Fig. 10.
The sensitivities of the couplings, varying one parameter at a time,
are summarized in Table II.
In each of these figures,
contours are shown for the muon mode alone and then for the combined
results of the $e$ and $\mu$ modes with both charge possibilities.
We also show contours for a reduced integrated luminosity of
300~fb$^{-1}$. For the $L_{9L}$ vs $L_{9R}$ plot we show contours
for both the electron and muon modes since there is a visible
difference for the two modes.
By combining the four
lepton modes the couplings can be measured to $\delta g_Z^1 =\pm 0.22$,
$\delta \kappa_Z = \pm 0.2$,
$\delta \kappa_\gamma =\pm 0.27$,  $\lambda_Z=\pm 0.18$,
$\lambda_\gamma = \pm 0.3$,
$\delta L_{9L}=\pm 55$, and $\delta L_{9R}= \sim\pm 300$.
If the results of the
four LEP experiments could be combined these results could be
reduced further.  Nevertheless, these limits are at the very least
an order of magnitude less sensitive than would be required to see
the effects of new physics through radiative corrections and are
comparable to the sensitivities that could be achieved at a high
luminosity Tevatron upgrade.  It is therefore unlikely, that new
physics will reveal itself at LEP200 through precision measurements
of the TGV's.

\subsubsection{$\sqrt{s}=500$~GeV}

For the $\sqrt{s}=500$~GeV NLC option we assume an integrated
luminosity of 50~fb$^{-1}$.    At 500~GeV the results are most
sensitive when we impose that the $W$'s are on mass-shell; ie.
$|M_{\ell\nu}-M_W|<10$~GeV and $|M_{q\bar{q}}-M_W|<10$~GeV.  This is
slightly more pronounced for the electron mode.  After these cuts
are imposed the electron and muon modes are essentially identical.
The 95\% confidence limits for the $g_Z^1-\kappa_Z$,
$\kappa_\gamma -\kappa_Z$, $\kappa_\gamma - \lambda_\gamma$, and
$\kappa_Z -\lambda_Z$ planes are shown in Fig. 11 and for the
$L_{9L}-L_{9R}$ planes in Fig. 12.
In each of these plots we show results for combining the four final
states and the $\mu^+$ mode alone.
We also show contours for
10~fb$^{-1}$ of integrated luminosity to show the effects of
reducing the collider luminosity.
The sensitivities, varying one at a time, are included in Table II.

Using the maximum likelihood analysis  we find
that $\kappa_\gamma$, $\kappa_Z$, $\lambda_\gamma$ and $\lambda_Z$
can be measured to better than $\pm 0.005$ and $g_Z^1$ to roughly $\pm
0.01$ at 95 \% C.L. including the invariant mass
cut and assuming 50~fb$^{-1}$ integrated luminosity.
If the integrated luminosity were reduced to 10~fb$^{-1}$ the bounds become
weaker by roughly a factor of two while combining all four modes
improves the single mode bounds by roughly a factor of two.
These measurements of $g_Z^1$, $\kappa_\gamma$, and $\kappa_Z$ should be
precise enough to probe loop radiative corrections to the
couplings.  On the other hand the measurements of $\lambda_\gamma$
and $\lambda_Z$ are still an order of magnitude too large to
see expected
deviations from tree level values due to radiative corrections.

The $L_{9L}-L_{9R}$ contours are shown in Fig. 12.
$L_{9L}$ can be measured to $\pm 3$  and $L_{9R}$ to $\sim \pm 8$
using a single mode and to $\pm 1.5$ and $\pm 4$ respectively
combining all four $\ell \nu q\bar{q}$ final states.

\subsubsection{$\sqrt{s}=1$~TeV}

For the 1~TeV NLC collider we assume an integrated luminosity of
200~fb$^{-1}$.  The 95\% C.L. sensitivity contours for the
$g_Z^1-\kappa_Z$ and $\kappa_\gamma -\kappa_Z$ are shown in Fig. 13.
These results were obtained by imposing that the $W$'s be
on mass shell;
$|M_{\ell\nu}-M_W|<10$~GeV and $|M_{q\bar{q}}-M_W|<10$~GeV.
We do not bother showing the $\lambda_V-\kappa_V$ contours since
they are uncorrelated (one parameter is least sensitive when the
other is taken to be equal to zero) so it is sufficient to give the
sensitivity when all other parameters are set to zero.  The
$L_{9L}-L_{9R}$ contours are shown in Fig 14.  With these collider
parameters $g_Z$ can be measured to about $\pm 0.005$ while
$\kappa_Z$ and $\kappa_\gamma$ can be measured to about $10^{-3}$.
These measurements will be sensitive enough to test the standard model at
the level of  radiative corrections.

\subsection{Initial State Polarization}

In the earlier discussion of angular distributions we pointed out
that reactions with different initial electron polarizations have
different dependences on anomalous couplings \cite{pankov95}.
In this section we
explore the consequences of this behavior.  We restrict our results
to the dimension 4 operators where deviations are most likely to
show up.

For the 500~GeV $e^+e^-$ collider we took 25~fb$^{-1}$ of integrated
luminosity per polarization.
We only include results for the four combined
lepton modes.  The 95\% C.L. contours for the $g_Z - \kappa_Z$,
$\kappa_\gamma -\kappa_Z$, and $L_{9L}-L_{9R}$ planes are shown in
Fig. 15.  Shown are contours for $e_L^- e^+$, $e_R^- e^+$, and
unpolarized  initial states. For the $g_Z-\kappa_Z$ plane
there is not much difference in the shape of the contour for the
different polarizations although the bounds improve slightly.  On
the other hand the different polarizations give much different
dependences for the  $\kappa_\gamma -\kappa_Z$
and $L_{9L}-L_{9R}$ contours.  In the $\kappa_\gamma -\kappa_Z$
plane the
right-handed electron polarizations give constraints orthogonal to
the left-handed polarizations and unpolarized results.  The
unpolarized contours are aligned along the $e_L^-$ contours which is
not too surprising considering that $\sigma (e^-_L)$ dominates
the right-handed contribution in the unpolarized cross-section. For
the $L_{9L}-L_{9R}$ plane the left and right handed polarizations
also give different dependences which would further constrain
$L_{9L}$.

For the 1~TeV $e^+e^-$ collider we took 100~fb$^{-1}$ of integrated
luminosity per polarization.  The sensitivities are shown in Fig.
16.  They are similar to, but more constraining, than the 500~GeV
case so we do not comment further.

More important  than the improvement in sensitivity is the
usefulness of polarization for
disentangling the nature of anomalous TGV's if deviations are
observed.

\subsection{Off Resonance Production}

In this section we explore the information potential available from
$\ell\nu q\bar{q}$ final states off the $W$ resonance.  Referring to
the invariant mass distributions shown in Fig. 4
one sees that there is considerable sensitivity to anomalous gauge
boson couplings when the fermion pairs do not originate from real
$W$ production.  We do not perform a rigorous analysis here but
demonstrate that there is considerable information in the
non-resonant production.   In particular we do not consider
possible backgrounds to non-resonant events and do not make any
effort to optimize our cuts to enhance deviations from SM results.
%We will suggest a strategy that makes optimal
%use of the information in each event.

We consider $\sqrt{s}=200$~GeV, 500~GeV, and 1~TeV for both the
$e\nu q\bar{q}$ and $\mu \nu q\bar{q}$ final states and include
initial state polarization when appropriate.
We based our results on the total cross-section upon imposing
the cuts $M_{ff'}<M_W-15$~GeV and $M_{ff'}>M_W+15$~GeV where
$M_{ff'}$ is the invariant mass of the final state fermion pairs and
$ff'$ stands for either $\ell \nu$ or $q\bar{q}$.  These give rise to a
large number of possibilities so we only present the
``best'' case when the four possible final states are combined for
each energy.

\subsubsection{$\sqrt{s}=200$~GeV}

For $\sqrt{s}=200$~GeV we only considered unpolarized initial state
electrons and positrons.  The results here should be taken with a
grain of salt due to the low number of events expected in these
kinematic regions.  For example, the standard model predicts,
for an integrated luminosity of
500~fb$^{-1}$ and combining all 4 final states, only 40 events when
either $M_{\ell\nu}<M_W-15$~GeV  or $M_{\ell\nu}>M_W+15$~GeV.  With
this warning, the optimum results occur for
$M_{\ell\nu}<M_W-15$~GeV and are given in Table III.  The results
are slightly weaker
for the case $M_{q\bar{q}}<M_W-15$~GeV. Although for a
specific case, sensitivities may differ between the $\mu$ and $e$
final states, they are generally quite similar.
The case $M_{ff'}>M_W+15$~GeV is not nearly as sensitive to anomalous
couplings except for $L_{9R}$ and $\kappa_\gamma$.
These results,
along with the previous ones which concentrated on the real $W$
production, indicate that the results are dominated by real $W$
production.
The off-resonance
results are roughly a factor of two to three weaker than those given
previously for real $W$ production and are not likely to contribute
much to bounds on TGV's at LEP200.

\subsubsection{$\sqrt{s}=500$~GeV}

For $\sqrt{s}=500$~GeV and combining the four final states,
the sensitivity is greatest when
$M_{\ell\nu}>M_W+15$~GeV except for a few cases.  The
results for $M_{q\bar{q}}$ cuts are slightly less sensitive.
Considering either the $e^\pm$ or $\mu^\pm$ final states separately
we find that for the $e^\pm$ final states the $M_{\ell\nu}>M_W+15$~GeV
case is more sensitive than the $M_{q\bar{q}}>M_W+15$~GeV case while
for the $\mu^\pm$ final states they are comparable.
With $M_{\ell\nu}>M_W+15$~GeV
the $e^\pm$ final states are  more sensitive to couplings involving
photons than the $\mu^\pm$
final states.  In both cases
when we take $M_{\ell\nu}>M_W+15$~GeV the cross-section is
dominated by the $q\bar{q}$ pair originating from an on-shell $W$.
This gives us the case of single-$W$ production which receives large
contributions from t-channel photon exchange and hence it is more
sensitive to the $WW\gamma$ coupling.

The NLC offers the possibility of initial electron polarization.  We
have included some representative results.  An important difference
between the two polarizations, which can be seen in Fig. 4, is that
the cross-section with left handed electrons is about an order of
magnitude larger than for right handed electrons.  At the same time
the right-handed cross-section is significantly more sensitive to anomalous
couplings than the left-handed cross-section.  There is therefore
a tradeoff between sensitivity and statistics so that in some cases
the bounds obtainable for the two polarizations are comparable.
The unpolarized results offer no improvement over the polarized
results since the right-handed
cross-section is overwhelmed by the left-handed contribution.  One
exception to these comments is when we consider $L_{9L}$
where the left-handed electrons are more constraining for
$L_{9L}$.

\subsubsection{$\sqrt{s}=1$~TeV}

The results at 1~TeV are qualitatively similar to those at 500~GeV
so we do not repeat the discussion of the previous section but only
point out the few points that differ.
Again, the highest sensitivity is for the constraint
$M_{\ell\nu}>M_W+15$~GeV.  The achievable bounds for this case are
included in Table III.  They are typically 4 to 5 times more
constraining than those obtainable at 500~GeV; less than 1\% for
$\kappa_\gamma$ and $\kappa_Z$ which is at the level of loop
contributions from new physics.

One interesting difference is that the muon mode for right-handed
initial electrons provides the most stringent constraints for many of
the TGV couplings.  As before, the electron final state
offers the best measurements of $\kappa_\gamma$.

\subsubsection{Comments on Off-Resonance Results}

{}From the above results it is clear that, although the constraints
that could be obtained from off-resonance production are not as
tight as those obtained from on-shell $W$ production,  there is
nevertheless considerable information contained in these events.
It appears to us that the method that makes optimal use of each
event is to calculate the probability of each event, irrespective of
where it appears in phase space, and compute a likelihood function
for the combined probabilities.  The only experimental cuts that
should be included are those that represent detector acceptance and
that are introduced to eliminate backgrounds.

\section{CONCLUSIONS}

We performed a detailed analysis of the measurement of tri-linear
gauge boson couplings in the process $e^+e^- \to \ell^\pm \nu q
\bar{q}$.  We included all tree level contributions to this final
state and included finite gauge boson width effects.  The off-shell $W$
contributions contribute from 20\% for the $\mu$ mode at LEP200 to
30\%
and 100\% for the electron mode at a 500~GeV and 1~TeV NLC
respectively (with the kinematic cuts we used).  Clearly,
the non-resonant contributions must be included to properly account
for the experimental situation.

To gauge the sensitivity of this process to anomalous gauge boson
couplings we used the $W$ decay distributions as a polarimeter to
distinguish the longitudinal $W$ modes, which are more sensitive to
anomalous coupings, from the transverse modes.  We implemented this
through the use of a quintic differential cross section, with each
angular variable divided into 4 bins, and then calculating the
likelihood function of non-standard model couplings as compared to
the standard model.  Using this approach we found that at LEP200
operating at 175~GeV and assuming integrated luminosity of
500~pb$^{-1}$, $g_1^Z$ $\kappa_Z$, $\kappa_\gamma$, $\lambda_Z$,
and $\lambda_\gamma$
could be measured to roughly $\pm 0.2$ and $L_{9L}$ and
$L_{9R}$ to $\pm 50$ and $\pm 400$ respectively. It is extremely
unlikely that measurements of this precision would reveal anomalous
couplings.
At a 500~GeV NLC with an integrated luminosity of
50~fb$^{-1}$, $g_1^Z$, $\kappa_V$ and $\lambda_V$,
could be measured to roughly $\pm 0.01$, $\pm 0.005$ and
$\pm 0.0025$ respectively and $L_{9L}$ and
$L_{9R}$ to $\pm 1$ and $\pm 4$ respectively. At the 1~TeV NLC with
200~fb$^{-1}$ the corresponding numbers are $\delta g_Z \sim \pm
0.05$, $\delta\kappa_{Z, \; \gamma} \sim \pm 10^{-3}$, $\delta
L_{9L} \sim \pm 0.5$ and $\delta L_{9R}\sim \pm 1$.
The 500~GeV NLC measurements are sensitive enough that they should
be sensitive to loop contributions to the TGV's while the 1~TeV will
be able to measure such effects.

We studied the sensitivity of the off-mass shell cross sections
to anomalous couplings by imposing kinematic cuts on the invariant
mass distributions of the outgoing fermion pairs.
A cursory analysis found that the off-resonance cross section is
relatively sensitive to anomalous couplings and that useful
information could be extracted from this region of phase space.

Although the inclusion of $W$ decays to fermions and the
non-resonant diagrams does not alter the precision to which the
TGV's can be measured they do change the cross sections and kinematic
distributions at the same level as radiative corrections and must be
taken into account for an accurate comparison between experiment and
theory.

The optimal strategy to maximize the information contained in each
event is to construct a likelihood function based on the four vector
of each of the outgoing fermions on an event by event basis, putting
them through a realistic detector simulation.  This would make the
best use of the information whether it be on the $W$ resonance or
not.  Kinematic cuts should only be introduced to reduce backgrounds.
Since the precision of these measurements is beyond the level of
loop induced radiative corrections it is crucial that radiative
corrections are well understood and included in event generators
used in the study of these processes.  Progress is being made along
these lines as exemplified by the Monte Carlo event generators;
{\tt EXCALIBUR} \cite{berends94b},
{\tt WHOPPER}, \cite{anlauf}
{\tt EEWW} \cite{fleischer94},
{\tt WWF} \cite{oldenborgh94},
and {\tt WWGENP} \cite{montagne95}.

\acknowledgments

The authors benefited greatly from many helpful conversations,
communications, and suggestions  during the course of this work
with Tim Barklow, Genevieve B\'elanger,
Pat Kalyniak, Dean Karlen, and Paul Madsen.
This research was supported in part by the Natural Sciences and
Engineering Research Council of Canada and Les Fonds FCAR du Quebec.

\figure{The Feynman diagrams contributing to the process
$e^+e^- \to \mu^+ \nu_\mu q \bar{q}'$.}

\figure{The Feynman diagrams that contribute to the process
$e^+e^- \to e^+ \nu_\mu q \bar{q}'$ in addition to those of fig 1.}

\figure{$\sigma(e^+e^-\to \mu^+\nu_\mu q\bar{q}')$ and
$\sigma(e^+e^-\to e^+\nu_e q\bar{q}')$ as a function of $\sqrt{s}$.
A $10^o$ cut away from the beam is imposed on charged final state
fermions and no cut on their energy.
In both cases the solid curve is the total cross section without any
cuts on the $\ell \nu$ and $q\bar{q}'$ invariant masses.  The dashed
curves are for the cut $|M_{q\bar{q}'}-M_W|<5$~GeV, the dotted
curves for $|M_{\ell\nu}-M_W|<5$~GeV and the dot-dashed curves for
both $|M_{q\bar{q}'}-M_W|<5$~GeV and  $|M_{\ell\nu}-M_W|<5$~GeV.}

\figure{Invariant mass distributions ($M_{\mu \nu}$, $M_{e\nu}$, and
$M_{q\bar{q}'}$) of final state fermions in the
processes $e^+ e^- \to \mu^+\nu_\mu q \bar{q}'$
and $e^+ e^- \to e^+\nu_e q \bar{q}'$ for $\sqrt{s}=500$~GeV.
Note the polarization of the
initial electron.  In all cases the solid line is the standard model
cross section, the long-dashed line is for $\kappa_Z=1.1$, the
dotted line is for $\lambda_Z=0.1$ and the dot-dashed line for
$\kappa_\gamma= 0.5$.}

\figure{The angular distribution of the outgoing $W^-$ with respect
to the incoming electron for $\sqrt{s}=200$~GeV, $\sqrt{s}=500$~GeV,
and $\sqrt{s}=1$~TeV.  The cross-section is given in units of
$R=4\pi\alpha^2/3s$. In all cases
the top solid line is for $e^-_L e^+ \to W_T W_T$,
the long-dashed line is for $e^-_L e^+ \to W_L W_T$,
the medium-dashed line is for $e^-_L e^+ \to W_L W_L$,
the short-dashed line is for the total of these three,
the dotted line is for $e^-_R e^+ \to W_T W_T$,
the dot-dashed line is for $e^-_R e^+ \to W_T W_L$,
the double dot-dashed line is for $e^-_R e^+ \to W_L W_L$,
and the bottom solid line is for the total of these last three.
Note that there is no bottom-solid line for $\sqrt{s}=500$~GeV.}

\figure{Angular distributions of $W$ for $\sqrt{s}=500$~GeV and
(a) $e_L^-$ and (b) $e_R^-$.
In both cases the solid line is the SM result, the dashed line is
for $\kappa_Z=1.1$, the dotted line for $\lambda_\gamma= -0.1$ and the
dot-dashed line for $\kappa_\gamma=0.5$. The distributions were obtained
from the full Monte Carlo by imposing the cut
$|M_{q\bar{q}}-M_W|<10$~GeV.}

\figure{Angular distributions of the outgoing quark with respect to the
$W^-$ direction in the $W$ rest frame.
In all cases the solid line is the SM result, the dashed line is
for $\kappa_Z=1.1$, the dotted line for $\lambda_Z= 0.1$ and the
dot-dashed line for $\lambda_\gamma=-0.1$. The distributions were obtained
from the full Monte Carlo by imposing the cut
$|M_{q\bar{q}}-M_W|<10$~GeV. }

\figure{Angle definitions used in our 5-dimensional angular
distribution analysis.  $\Theta$ is the $W$
scattering angle, $\theta_{qq}$ and $\theta_{\ell\nu}$ are the decay
angles in the $W$ rest frames and $\phi_{qq}$ and $\phi_{\ell\nu}$ are
the azimuthal angles, again in the $W$ rest frames.}

\figure{95\% C.L. contours for sensitivity to anomalous couplings for
$\sqrt{s}=175$~GeV.  In all cases the inner solid contour is obtained from
combining all 4 lepton charge states for L=500~pb$^{-1}$,
the heavy outer solid line is for the $\mu^+$ mode alone for L=500~pb$^{-1}$,
and the dotted contour is for the reduced luminosity case of
L=300~pb$^{-1}$ with all 4 modes combined.}

\figure{95\% C.L. contours  for sensitivity to $L_{9L}$ and $L_{9R}$ for
$\sqrt{s}=175$~GeV.  (a) The heavy solid line is for the $\mu^+$ mode, the
dotted line is for the $e^+$ mode and the inner solid line is for
combining all four lepton charge states; all for  L=500~pb$^{-1}$.
(b) Both curves are from combining all four lepton charge states. The solid
line is for  L=500~pb$^{-1}$ and the dotted line is for
L=300~pb$^{-1}$.}

\figure{95\% C.L. contours for sensitivity to anomalous couplings for
$\sqrt{s}=500$~GeV.  In all cases the inner solid contour is obtained from
combining all 4 lepton charge states for L=50~fb$^{-1}$,
the heavy solid line is for the $\mu^+$ mode alone for L=50~fb$^{-1}$,
and the dotted contour is for the reduced luminosity case of
L=10~fb$^{-1}$ with all 4 modes combined.
These results were obtained by imposing that the $W$'s are
on mass shell; $|M_{\ell\nu}-M_W|<10$~GeV and $|M_{q\bar{q}}-M_W|<10$~GeV.}

\figure{95\% C.L. contours  for sensitivity to $L_{9L}$ and $L_{9R}$ for
$\sqrt{s}=500$~GeV.
The inner solid line is obtained by combining all four lepton charge states
and is for L=50~fb$^{-1}$,  the heavy solid line is for all 4 modes and
L=10~fb$^{-1}$
and the dotted line is for the $\mu^+$ mode alone for  L=50~fb$^{-1}$.
These results were obtained by imposing that the $W$'s are
on mass shell; $|M_{\ell\nu}-M_W|<10$~GeV and $|M_{q\bar{q}}-M_W|<10$~GeV.}

\figure{95\% C.L. contours for sensitivity to anomalous couplings for
$\sqrt{s}=1$~TeV.  In both cases the inner solid contour is obtained from
combining all 4 lepton charge states for L=200~fb$^{-1}$ and
the dotted contour is for the reduced luminosity case of
L=50~fb$^{-1}$ with all 4 modes combined.  The $\mu^+$ contour
for L=200~fb$^{-1}$ lies on top of the dotted curves.
These results were obtained by imposing that the $W$'s are
on mass shell; $|M_{\ell\nu}-M_W|<10$~GeV and $|M_{q\bar{q}}-M_W|<10$~GeV.}

\figure{95\% C.L. contours  for sensitivity to $L_{9L}$ and $L_{9R}$ for
$\sqrt{s}=1$~TeV.
The inner solid line is obtained by combining all four lepton charge states
and is for L=200~fb$^{-1}$  and the dotted line is for all 4 modes and
L=50~fb$^{-1}$.  The $\mu^+$ mode alone for  L=200~fb$^{-1}$ sits on top
of the dotted contour.
These results were obtained by imposing that the $W$'s are
on mass shell; $|M_{\ell\nu}-M_W|<10$~GeV and $|M_{q\bar{q}}-M_W|<10$~GeV.}

\figure{95\% C.L. contours for sensitivity to anomalous couplings for
polarized initial state electrons for
$\sqrt{s}=500$~GeV and  L=25~fb$^{-1}$ per polarization and
combining all four lepton charge states.
In all cases the solid curves are for $e^-_{L}$, the dashed curves for
$e^-_{R}$, and the heavy solid curve for unpolarized electrons (for
a total of L=50~fb$^{-1}$).}

\figure{95\% C.L. contours for sensitivity to anomalous couplings for
polarized initial state electrons for
$\sqrt{s}=1$~TeV and  L=100~fb$^{-1}$ per polarization
and combining all four lepton charge states.
In all cases the solid curves are for $e^-_{L}$, the dashed curves for
$e^-_{R}$, and the heavy solid curve for unpolarized electrons
(for a total of L=200~fb$^{-1}$).}

\newpage

\begin{table}
\caption{Cross-sections for $e^+e^- \to \mu^+\nu_\mu q\bar{q}'$ and
$e^+e^- \to e^+\nu_e q\bar{q}'$ including cuts on the invariant
masses of the outgoing fermion pairs, $M_{\ell\nu}$ and $M_{q\bar{q}'}$.
The cross-sections are given in pb.}
\label{table1}
\begin{tabular}{llllll}
$\sqrt{s}$ & $\ell$ & no cut & $|M_{q\bar{q}}-M_W|<5$~GeV &
	$| M_{\ell\nu}-M_W|<5$~GeV & both cuts \\
(GeV) &   &  &  &  \\
\tableline
175 & $\mu$ & 1.10 & 1.00 & 1.00 & 0.91 \\
	& $e$ & 1.15 & 1.04 & 1.01 & 0.91 \\
500 & $\mu$ & 0.39 & 0.34 & 0.34 & 0.29 \\
	& $e$ & 0.62 & 0.53 & 0.34 & 0.29 \\
1000 & $\mu$ & 0.077 & 0.063 & 0.064 & 0.052 \\
	& $e$ & 0.44 & 0.39 & 0.064 & 0.052 \\
\end{tabular}
\end{table}

\begin{table}
\caption{Sensitivities to anomalous couplings for the various
parameters varying one parameter at a time.  The values are obtained
by combining the four lepton modes ($e^-$, $e^+$, $\mu^-$, and
$\mu^+$) and two generations of light quarks ($ud$, $cs$).
%We
%imposed the cuts $|M_{\ell\nu (q\bar{q})} -M_W|<10$~GeV on the
%invariant masses of the $\ell\nu$ $(q\bar{q})$ pairs.
The results are  95\% confidence level limits.}
\label{thetable}
\begin{tabular}{lccccccc}
mode & $L_{9L}$ & $L_{9R}$ & $\delta g_1^Z$ & $\delta\kappa_Z$ &
	$\delta\kappa_\gamma$ & $\lambda_Z$ & $\lambda_\gamma$ \\
\tableline
\multicolumn{8}{c}{$\sqrt{s}=175$~GeV, L=500~pb$^{-1}$, no cuts on
$M_{\ell\nu (q\bar{q})}$ }\\
\tableline
$\mu$ & $\pm 110$ & $^{+920}_{-420}$ & $^{+0.45}_{-0.44}$ &
	$^{+0.39}_{-0.38}$ & $^{+0.58}_{-0.48}$ & $^{+0.36}_{-0.35}$
	& $^{+0.61}_{-0.50}$ \\
$e$ & $^{+120}_{-110}$ & $^{+620}_{-440}$ & $^{+0.44}_{-0.43}$ &
	$^{+0.40}_{-0.38}$ & $^{+0.58}_{-0.52}$ & $^{+0.37}_{-0.35}$
	& $^{+0.62}_{-0.49}$ \\
combined & $\pm 55$ & $^{+330}_{-230}$ & $\pm 0.22$ &
	$^{+0.19}_{-0.20}$ & $^{+0.27}_{-0.26}$ & $\pm 0.18$
	& $^{+0.29}_{-0.26}$ \\
\tableline
\multicolumn{8}{c}{$\sqrt{s}=500$~GeV, L=50~fb$^{-1}$,
$|M_{\ell\nu (q\bar{q})}- M_W|<10$~GeV }\\
\tableline
$\mu$ & $^{+2.2}_{-2.1}$ & $^{+4.6}_{-4.2}$ & $\pm 0.020$ &
	$\pm 0.007$ & $\pm 0.005$ & $\pm 0.005$
	& $\pm 0.006$ \\
$e$ & $^{+2.2}_{-2.1}$ & $^{+4.6}_{-4.2}$ & $^{+0.019}_{-0.020}$ &
	$\pm 0.007$ & $\pm 0.005$ & $\pm 0.005$
	& $\pm 0.006$ \\
combined & $^{+1.1}_{-1.0}$ & $^{+2.2}_{-2.1}$ & $\pm 0.0095$ &
	$\pm 0.0035$ & $\pm 0.0025$ & $\pm 0.0025$
	& $\pm 0.0025$ \\
\tableline
\multicolumn{8}{c}{$\sqrt{s}=1$~TeV, L=200~fb$^{-1}$,
$|M_{\ell\nu (q\bar{q})}- M_W|<10$~GeV }\\
\tableline
$\mu$ & $^{+0.61}_{-0.62}$ & $^{+1.3}_{-1.1}$ & $\pm 0.01$ &
	$\pm 0.002$ & $\pm 0.001$ & $\pm 0.002$
	& $\pm 0.002$ \\
$e$ & $^{+0.61}_{-0.62}$ & $^{+1.3}_{-1.1}$ & $\pm 0.01$ &
	$\pm 0.002$ & $\pm 0.001$ & $\pm 0.002$
	& $\pm 0.002$ \\
combined & $\pm 0.28$ & $^{+0.62}_{-0.56}$ & $\pm 0.0054$ &
	$\pm 0.001$ & $\pm 0.0006$ & $\pm 0.0008$
	& $\pm 0.0008$ \\
\end{tabular}
\end{table}

\begin{table}
\caption{Sensitivities to anomalous couplings based on off-resonance
cross sections varying one parameter at a time.  The values are obtained
by combining the four lepton modes ($e^-$, $e^+$, $\mu^-$, and
$\mu^+$) and two generations of light quarks ($ud$, $cs$).
The results
are  95\% confidence level limits. A dash signifies that the bound
is significantly weaker than the others.}
\label{thetable}
\begin{tabular}{ccccccccc}
Initial State & cut & $L_{9L}$ & $L_{9R}$ & $\delta g_{1Z}$
	& $\delta\kappa_Z$ &
	$\delta\kappa_\gamma$ & $\lambda_Z$ & $\lambda_\gamma$ \\
\tableline
\multicolumn{9}{c}{$\sqrt{s}=200$~GeV, L=500~pb$^{-1}$ }\\
\tableline
$e^-$ & $M_{\ell\nu}<M_W -15$~GeV & $_{-320}^{+400}$ & $_{-780}^{+520}$ &
$_{-1.1}^{+1.5}$ & $_{-0.9}^{+1.1}$ & $_{-1.0}^{+0.8}$ &
$_{-1.0}^{+1.0}$ & $_{-0.7}^{+1.1}$ \\
\tableline
\multicolumn{9}{c}{$\sqrt{s}=500$~GeV, L=50~fb$^{-1}$ }\\
\tableline
$e_L^-$ & $M_{\ell\nu}>M_W +15$~GeV & $_{-7.2}^{+6.9}$ &
$_{-9.2}^{+8.9}$ & $_{-0.18}^{+0.13}$
	& $_{-0.074}^{+0.056}$ & $_{-0.012}^{+0.012}$ &
$_{-0.034}^{+0.026}$ & $_{-0.041}^{+0.026}$ \\
$e_R^-$ & $M_{\ell\nu}>M_W +15$~GeV & $_{---}^{---}$ & $_{---}^{---}$ &
$_{-0.24}^{+0.08}$ & $_{-0.10}^{+0.03}$ & $_{-0.018}^{---}$ &
$_{-0.033}^{+0.023}$ & $_{-0.023}^{+0.034}$ \\
$e^-$ & $M_{\ell\nu}>M_W +15$~GeV & $_{-10}^{+10}$ & $_{-13}^{+13}$ &
$_{-0.21}^{+0.15}$ & $_{-0.10}^{+0.07}$ & $_{-0.017}^{+0.016}$ &
$_{-0.038}^{+0.031}$ & $_{-0.043}^{+0.033}$ \\
\tableline
\multicolumn{9}{c}{$\sqrt{s}=1$~TeV, L=200~fb$^{-1}$}\\
\tableline
$e_L^-$ & $M_{\ell\nu}>M_W +15$~GeV & $_{-2.5}^{+2.4}$ &
$_{-3.9}^{+3.8}$  &
$_{-0.085}^{+0.038}$ & $_{-0.014}^{+0.013}$ & $_{-0.005}^{+0.005}$ &
$_{-0.002}^{+0.002}$ & $_{-0.003}^{+0.003}$ \\
$e_R^-$ & $M_{\ell\nu}>M_W +15$~GeV & $_{---}^{---}$ &
$_{-5.9}^{---}$ & $_{-0.102}^{+0.062}$ & $_{---}^{+0.010}$ &
$_{-0.010}^{---}$ & $_{-0.007}^{+0.007}$ & $_{-0.007}^{+0.007}$ \\
$e^-$ & $M_{\ell\nu}>M_W +15$~GeV & $_{-3.5}^{+3.4}$ &
$_{-5.6}^{+5.4}$ & $_{-0.096}^{+0.047}$
	& $_{-0.021}^{+0.018}$ & $_{-0.007}^{+0.007}$ &
$_{-0.003}^{+0.003}$ & $_{-0.003}^{+0.003}$ \\
%\tableline
\end{tabular}
\end{table}


\begin{references}
\bibitem{lep}
	D. Schaile, {\sl Proceedings of the XXVII Int. Conf. on High
	Energy Physics}, eds. P.J. Bussey, I.G. Knowles,
	Glasgow, UK, 20-27 July 1994 (Inst. of Physics Publishing,
	1995) p. 27.
\bibitem{hagiwara87}
	K. Hagiwara {\it et al}, Nucl. Phys. {\bf B282}, 253 (1987).
\bibitem{lep200}
	D. Zeppenfeld, Phys. Let. {\bf 183B}, 380 (1987);
	D. Treille {\it et al}, Proceedings of the ECFA Workshop on LEP 200,
	ed. A. B\"ohm and W. Hoogland, Aachen (1986), CERN 87-08, vol.2, p.414.
	D.A. Dicus, K. Kallianpur, Phys. Rev. {\bf D32}, 35 (1985);
	M.J. Duncan, G.L. Kane, Phys. Rev. Lett. {\bf 55}, 773 (1985);
        E.N.Argyres\ and C.G.Papadopoulos, Phys. Lett. {\bf B263}, 298(1991).
\bibitem{kane89}
	G.Kane, J. Vidal, C.P. Yuan, Phys. Rev. {\bf D39}, 2617 (1990),
	and references therein.
\bibitem{nlcphysics}
	Some recent reviews on physics at high energy $e^+e^-$
	colliders are:
	M. Peskin, Proceedings of the 1987 SLAC Summer Institute on
	Particle Physics, SLAC-PUB-4601 (1988);
	D. Treille, Proceedings of the Workshop on Physics and
	Experiments with Linear $e^+e^-$ Colliders,
	Waikoloa Hawaii, April 1993 (World Scientific; in press);
	A. Djouadi and P.M. Zerwas, Proceedings of {\sl Beyond the Standard
	Model III}, ed. S. Godfrey and P. Kalyniak, Ottawa Canada, June 1992,
	(World Scientific, Singapore, 1993) p. 204.
\bibitem{miyamoto}
	A. Miyamoto, Proceedings of Workshop on Physics and Experiments
	with Linear $e^+e^-$ Colliders, Waikoloa, Hawaii, ed. F.A. Harris,
	S.L. Olsen, S. Pakvasa, and X. Tata, (World Scientific, Singapore,
	1994) p 141.
\bibitem{nlc}
	B. Wiik, Proceedings of the Workshop on Physics and
	Experiments with Linear $e^+e^-$ Colliders,
	Waikoloa Hawaii, ed. F.A. Harris,
	S.L. Olsen, S. Pakvasa, and X. Tata, (World Scientific, Singapore,
	1994).
\bibitem{JLC}
	Proceedings of the First Workshop on Japan Linear Collider (JLC),
	KEK, Oct 24-25, 1989, KEK Report 90-2 (1990);
	Proceedings of the Second Workshop on Japan Linear Collider,
	ed. S. Kawabata, Nov. 6-8, 1990 KEK Proceedings 91-10 Nov. 1991.
\bibitem{NLC2}
	C. Ahn {\it et al.},
	{\sl Opportunities and Requirments for Experimentation
	at a Very High Energy $e^+e^-$ Collider}, SLAC Report SLAC-0329 (1988);
\bibitem{CLIC}
	Proceedings of the Workshop on Physics at Future Accelerators,
	CERN Yellow Report 87-07 (1987).
\bibitem{aihara95}
	For a recent comprehensive review of trilinear gauge boson couplings see:
	H. Aihara {\it et al.}, To appear in {\sl Electroweak
	Symmetry Breaking and Beyond the Standard Model}, eds. T. Barklow,
	S. Dawson, H. Haber and J. Siegrist (World Scientific),
	{\tt hep-ph/9503425}.
\bibitem{tgvreviews}
	For further recent reviews on TGV's see:
	I. Hinchliffe, To appear in the proceedings of the
	International Symposium on Vector Boson Self Interactions,
	UCLA, Feb 1-3, 1995, {\tt hep-ph/9504206};
	F. Boudjema, To appear in the proceedings of {\sl Beyond the
	Standard Model IV}, Lake Tahoe, California, Dec. 12-16, 1994,
	{\tt hep-ph/9504409};
\bibitem{burgess94}
	C.P. Burgess, S. Godfrey, H. K\"onig, D. London, and I. Maksymyk,
	Phys. Rev. {\bf D49}, 6115 (1994).
\bibitem{dawson95}
	S. Dawson and G. Valencia, Nucl. Phys. {\bf B439}, 3 (1995).
\bibitem{hagiwara93}
	K. Hagiwara, S. Ishihara, R. Szalapski, and D. Zeppnefeld,
	Phys. Lett. {\bf B283}, 353 (1992); Phys. Rev. {\bf D48}, 2182 (1993).
\bibitem{tevatron}
	F. Abe {\it et al.} (CDF Collaboration), Phys. Rev. Lett.
	{\bf 74}, 1936 (1995);
	J. Ellison {\it et al.} (D0 Collaboration), Proceedings of the
	{\sl DPF'94 Conference}, Albuquerque, NM, August 1994;
	H. Aihara, to appear in the Proceedings of the
	{\sl International Symposium on Vector Boson
	Self-Interactions}, UCLA, February 1995;
	T.A. Fuess (CDF Collaboration), Proceedings of the
	{\sl DPF'94 Conference}, Albuquerque, NM, August 1994;
	F. Abe {\it et al.}, (CDF Collaboration), Proceedings of the
	{\sl 27th International Conference on High Energy Physics},
	Glasgow, Scotland, July 20-27, 1994;
	F. Abe {\it et al.}, (CDF Collaboration),
	FERMILAB-Pub-95/036-E, submitted to Phys. Rev. Lett.;
	S. Abachi {\it et al.}, (D0 Collaboration),
	FERMILAB-Pub-95/101-E, {\tt hep-ex/9505007}.
\bibitem{lhc}
	U. Baur and D. Zeppenfeld, Nucl. Phys. {\bf B308}, 127 (1988);
	see also H. Aihara {\it et al.} Ref. \cite{aihara95}.
\bibitem{yehudai}
	E. Yehudai, Phys. Rev. {\bf D 41}, 33 (1990); {\bf D44}, 3434 (1991);
	S.Y. Choi and F. Schrempp, Phys. Lett. {\bf 272B}, 149 (1991);
	I.F. Ginzburg, G.L. Kotkin, S.L. Panfil and V.G. Serbo,
	Nucl. Phys. {\bf B228}, 285 (1983)
\bibitem{godfrey92}
	S. Godfrey and K.A. Peterson, Carleton University report
	OCIP/C-92-7, {\tt hep-ph/9302319}.
\bibitem{couture}
	G. Couture, S. Godfrey, P. Kalyniak Phys. Rev. {\bf D39}, 3239 (1989);
	Phys. Rev. {\bf D42},1841 (1990);
	Phys. Lett. {\bf B218}, 361 (1989).
\bibitem{egamma}
	M. Raidal, University of Helsinki report HU-SEFT R 1994-16,
	{\tt hep-ph/9411243};
	K. Cheung, S. Dawson, T. Han, and G. Valencia,
	Phys. Rev. {\bf D51}, 5 (1995) ({\tt hep-ph/9403358});
	A. Queijeiro, Phys. Lett. {\bf B193}, 354 (1987);
	K.J. Abraham and C.S. Kim, Phys. Lett. {\bf B301}, 430 (1993);
	S.J. Brodsky, T.G. Rizzo, and I. Schmidt, SLAC-PUB-95-6904,
	{\tt hep-ph/9505441}.
\bibitem{cuypers93}
	D. Choudhury and F. Cuypers, {\tt hep-ph/9312308}.
\bibitem{choi91}
        G. Couture and G. B\'elanger, Phys. Rev. {\bf D49}, 5720 (1994).
\bibitem{couture92}
	G. Couture, S. Godfrey, and R. Lewis, Phys. Rev. {\bf D45}, 777 (1992);
	G. Couture and S. Godfrey, Phys. Rev. {\bf D49}, 5709(1994).
\bibitem{ambrosanio92}
	S. Ambrosanio and B. Mele, Nucl. Phys. {\bf B374}, 3 (1992).
\bibitem{hagiwara91}
	K. Hagiwara {\it et al.}, Nucl. Phys. {\bf B365}, 544 (1991).
\bibitem{couture94}
	G. Couture and S. Godfrey, Phys. Rev. {\bf D50}, 5607(1994);
	K.J. Abraham, J. Kalinowski, and P. \'Sciepko, Proceedings
	of the XVII Warsaw Symposium on Elementary Particle Physics,
	Kazimierz, May 23-27, 1994, {\tt hep-ph/9407223}.
\bibitem{eptog}
	G.V. Borisov, V.N. Larin, and F.F. Tikhonin,
	Zeit. Phys. {\bf C41}, 287 (1988).
\bibitem{barklow92}
	T. Barklow, {\sl Proceedings of the 1st Workshop on Physics
	with Linear Colliders}, Saariselka, Finland, Sept. 9-14 (1992);
	T. Barklow, {\sl Proceedings of the DPF'94 Conference: 1994
	Meeting of the Division of Particls and Fields of the APS},
	Albuquerque, NM, Aug2-6, 1994.
\bibitem{papa95}
	C.G. Papadopoulos, CERN Report CERN-TH/95-46, {\tt hep-ph/9503276}.
\bibitem{sekulin}
	R.L. Sekulin, Phys. Lett. {\bf B338}, 369 (1994).
\bibitem{eptoww}
	M. Bilenky, J.L. Kneur, F.M. Renard and D. Schildknecht,
	Nucl. Phys. {\bf B409}, 22 (1993);
	G. Gounaris, J.L. Kneur, J. Layssac, G. Moultaka, F.M.
	Renard, and D. Schildknecht, {\sl $e^+e^-$ Collisions at
	500~GeV, the Physics. Potential}, ed. P. Zerwas, DESY (Hamburg, 1991);
	K.J.F. Gaemers and M.R. van Velzen, Z. Phys. {\bf C43}, 103 (1989);
	M. Diehl and O. Nachtmann, Z. Phys. {\bf C62}, 397 (1994).
\bibitem{pankov95}
	A.A. Pankov and N. Paver, Phys. Lett. {\bf B324}, 224 (1994);
	{\bf B346}, 115 (1995);
	A.A. Likhoded, A.A. Pankov, N. Paver, M.V. Shevlyagin and
	O.P. Yushchenko, UTS-DFT-93-22 (1993).
\bibitem{beenakker94}
	For a detailed review see W. Beenakker and A. Denner,
	International Journal of Modern Physics {\bf A9}, 4837(1994).
\bibitem{kalyniak93}
	P. Kalyniak, P. Madsen, N. Sinha, and R. Sinha,
	Phys. Rev. {\bf D48}, 5081 (1993).
\bibitem{gintner95}
	M. Gintner and S. Godfrey, Phys. Lett. {\bf B}, (in press),
	{\tt hep-ph/9510324}.
\bibitem{kurihara94}
	Y. Kurihara, D. Perret-Gallix, and Y. Shimizu, Phys. Lett.
	{\bf B349}, 367 (1995) ({\tt hep-ph/9412215});
	M. Nowakowski and A. Pilaftsis, Z. Phys. {\bf C};
	A. Aeppli, F. Cuypers and G. J van Oldenborgh,
	Phys. Lett. {\bf B314}, 413 (1993).
\bibitem{fourfermions}
	F.A. Berends, R. Pittau, and R. Kleiss, Nucl. Phys.
	{\bf B424}, 308(1994);
	G. Montagna, O. Nicrosini, G. Passarino, and F. Piccinini,
	CERN Report CERN-TH.7497/94 (1994, unpublished);
	D. Bardin, A. Leike, and T. Riemann,
	CERN Report CERN-TH.7478/94 (1994, unpublished), {\tt hep-ph/9410361};
	G. Montagna, O. Nicrosini, G. Passarino, and F. Piccinini,
	CERN Report CERN-TH.7497/94 (1994; unpublished);
	J. Fujimoto {\it et al.}, KEP Report KEK 94-46;
	T. Ishikawa {\it et al.}, Proceedings of the {\sl VIIth
	Workshop on High Energy Physics and Quantum Field Theory},
	Sotchi, Russia, Oct. 7-14 1992.
\bibitem{singlew}
	H. Iwasaki, Int. Journal of Mod. Phys. {\bf A7}, 3291(1992);
	H. Neufeld, Z. Phys. {\bf C17}, 145 (1983);
	E.N. Argyres and C.G. Papadopoulos, Phys. Lett. {\bf 263B}, 298 (1991);
	O. Philipsen, Z. Phys. {\bf C54}, 643 (1992);
	G. Couture and J.N. Ng, Z. Phys. {\bf C32}, 579 (1986);
	J.C. Romao and P. Nogueira, Z. Phys. {\bf C42}, 263 (1989);
	O. Cheyette, Phys. Lett. {\bf 137B}, 431 (1984);
	C.G. Papadopoulos, Phys. Lett. {\bf B333}, 202 (1994);
	E. Gabrielli, Mod.  Phys. Lett. {\bf A1}, 465 (1986).
\bibitem{emrc}
	O. Nicrosini and L. Trendadue, 	Nucl. Phys. {\bf B318}, 1 (1989);
	L. Trentadue {\it et al.}, in : Z Physics at LEP1, ed. G.
	Altarelli, CERN Yellow report CERN 89-08 (CERN, Geneva, 1989) p129;
	D. Bardin, M. Bilenky, A. Olchevski, and T. Riemann,
	Phys. Lett. {\bf B308}, 403 (1993);
\bibitem{berends94b}
	F.A. Berends, R. Pittau, and R. Kleiss, Nucl. Phys.
	{\bf B426}, 344(1994); 	{\tt hep-ph/9409326};
	R. Pittau, {\tt hep-ph/9406233}.
\bibitem{beenakker91}
	W. Beenakker, K. Kolodziej, and T. Sack,
	Phys. Lett. {\bf B258}, 469 (1991);
	W. Beenakker and A. Denner, DESY 94-051;
	W. Beenakker, F.A. Berends, and T. Sack,
	Nucl. Phys. {\bf B367}, 287 (1991).
\bibitem{fleischer94}
	J. Fleischer, F. Jegerlehner and M. Zralek,
		Z. Phys. {\bf C42}, 409 (1989);
	J. Fleischer, F. Jegerlehner, K. Koodziej, and G.J. van Oldenborgh,
	preprint PSI-PR-94-16 (1994; unpublished), {\tt hep-ph/9405380}.
\bibitem{oldenborgh94}
	G. Jan van Oldenborgh, P.J. Franzini, and A. Borrelli,
	{\tt hep-ph/9402298}.
\bibitem{bardin94}
	D. Bardin {\it et al.}, CERN Report CERN-TH.7295/94,
	{\tt hep-ph/9406340};
	J. Fujimoto {\it et al.}, {\tt hep-ph/9407308};
	G. Montagna {\it et al.}, CERN Report CERN-TH.7497/94,
	{\tt hep-ph/9411332};
	Yu.L. Dokshitzer, V.A. Khoze, L.H. Orr, and W.J. Stirling,
	Phys. Lett. {\bf B313}, 171 (1993).
\bibitem{gintner95b}
	M. Gintner, S. Godfrey and G. Couture, in preparation.
\bibitem{boudjema}
	For recent reviews of effective Lagrangians see:
	M.B. Einhorn Proceedings of the {\it Workshop on Physics
	and Experiments with Linear $e^+e^-$ Colliders},eds. F.A.
	Harris {\it et al.}, Waikoloa
	Hawaii, April 26-30, 1993 (World Scientific, 1994) p. 122;
	F. Boudjema, {\it ibid}, p.712.
\bibitem{bagger93}
J. Bagger, S. Dawson, and G. Valencia, Nucl. Phys. {\bf B399}, 364 (1993).
\bibitem{gaemers79}
	K. Gaemers and G. Gounaris, Z. Phys. {\bf C1}, 259 (1979).
\bibitem{miscvertex}
	K.-i Hikasa Phys. Rev. {\bf D33}, 3203 (1986);
	K. Hagiwara {\it et al}, Nucl. Phys. {\bf B282}, 253 (1987).
\bibitem{cp}
	W.J. Marciano, A. Queijeiro, Phys. Rev. {\bf D 33}, 3449 (1986);
	F. Boudjema, K. Hagiwara, C. Hamzaoui, and K. Numata, Phys. Rev.
	{\bf D43}, 2223 (1991).
\bibitem{baur88}
	U. Baur and D. Zeppenfeld, Nucl. Phys. {\bf B308}, 127 (1988).
\bibitem{smloops}
	W.A. Bardeen, R. Gastmans, and B. Lautrup, Nucl. Phys.
	{\bf B46}, 319 (1972);
	E.N. Argyres {\it et al.,} Nucl. Phys. {\bf B391}, 23 (1993);
	J. Papavassiliou and K. Philoppides,
	Phys. Rev. {\bf D48}, 4255 (1993);
	G. Couture and J.N. Ng, Z. Phys. {\bf C35}, 65 (1987);
	G. Couture {\it et al.}, Phys. Rev. {\bf D38}, 860 (1988).
\bibitem{kuroda87}
	M. Kuroda, F.M. Renard, and D. Schildnecht,
	Phys. Lett. {\bf B183}, 366 (1987).
\bibitem{nonlinear}
	A. Longhitano, Nucl. Phys. {\bf B188}, 118 (1981);
	T. Appelquist and C. Bernard, Phys. Rev. {\bf D22}, 200 (1980);
	B. Holdom, Phys. Lett. {\bf B258}, 156 (1991);
	A. Falk, M. Luke, and E. Simmons, Nucl. Phys. {\bf B365}, 523 (1991);
	T. Appelquist and G.H. Wu, Phys. Rev. {\bf D48}, 3235 (1993);
	see also J.F. Donoghue, E. Golowich, and B.R. Holstein,
	{\sl Dynamics of the Standard Model},
	(Cambridge University Press, 1994).
\bibitem{linear}
	W. Buchm\"uller and D. Wyler, Nucl. Phys. {\bf B268}, 621 (1986);
	A. De R\'ujula, M.B. Gavela, P. Hernandez and E. Mass\'o,
	Nucl. Phys. {\bf B384}, 3 (1992)
\bibitem{zeppenfeld92}
	D. Zeppenfeld, J.A.M. Vermaseren and U. Baur,
	Nucl. Phys. {\bf B375}, 3 (1992).
\bibitem{baur95}
	U. Baur and D. Zeppenfeld, MAD/PH/878.
\bibitem{calkul}
	R.~Kleiss and W.~J.~Stirling, Nucl. Phys. {\bf B262}, 235 (1985);
	Z.Xu D.-H.Zhang L. Chang Nucl. Phys. {\bf B291}, 392 (1987).
\bibitem{monte}
	See for example V. Barger and R. Phillips, {\sl Collider Physics},
	(Addison-Wesley Publishing Company, 1987).
\bibitem{peskin88}
	M. Peskin, Proceedings of the 1987 SLAC Summer Institute on
	Particle Physics, SLAC-PUB-4601 (1988).
\bibitem{burke91}
	D.L. Burke, Proceedings of the 1990 SLAC Summer Institute on
	Particle Physics, SLAC-PUB-5418 (1991).
\bibitem{anlauf}
	H. Anlauf {\it et al.}
	Proceedings of Workshop on Physics and Experiments
	with Linear $e^+e^-$ Colliders, Waikoloa, Hawaii, ed. F.A. Harris,
	S.L. Olsen, S. Pakvasa, and X. Tata, (World Scientific, Singapore,
	1994) p 708.
\bibitem{montagne95}
	G. Montagne, O. Nicrosini, F. Piccinini,
	Comp. Phys. Commun. {\bf 90}, 141 (1995).
\end{references}
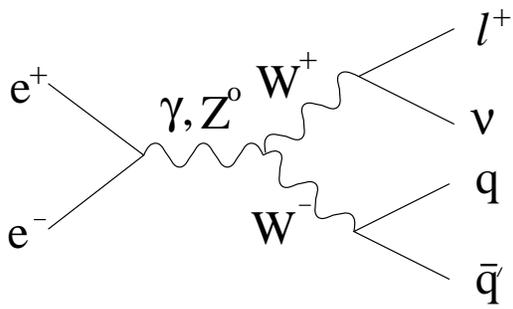
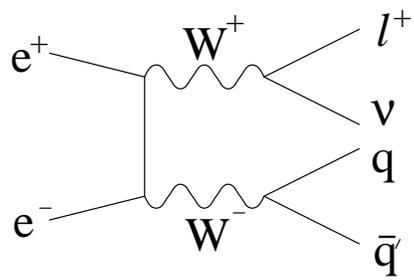
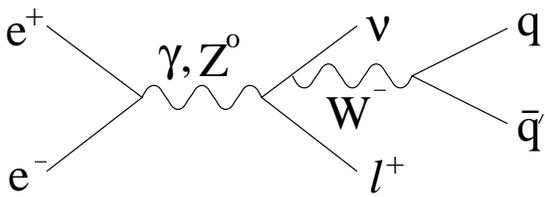
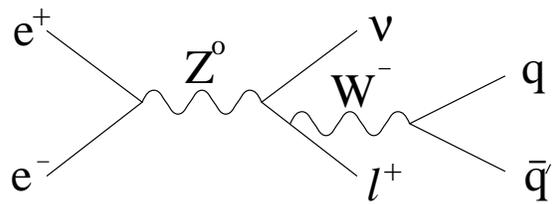
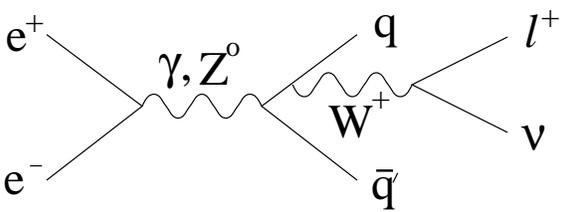
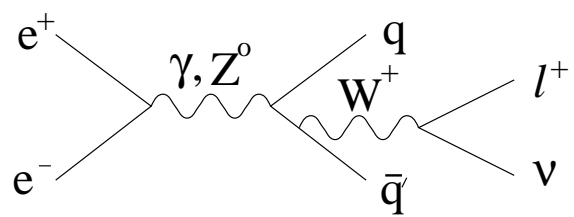
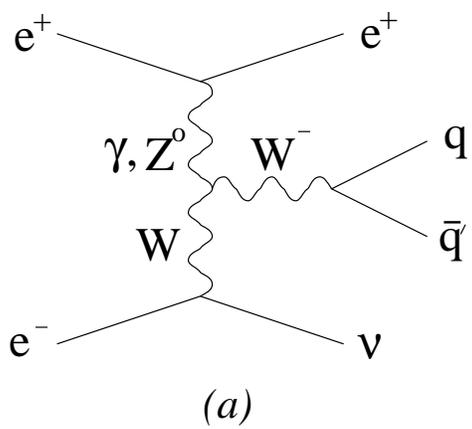
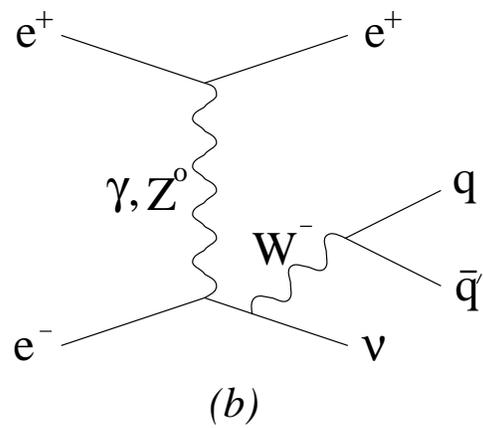
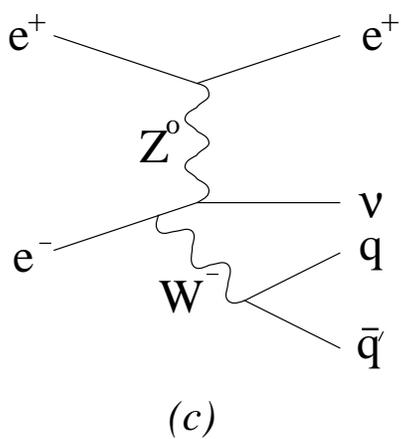
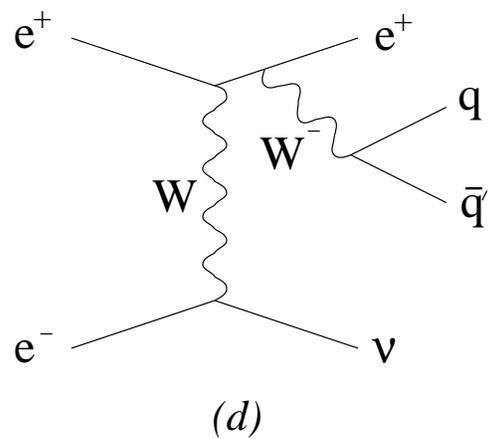
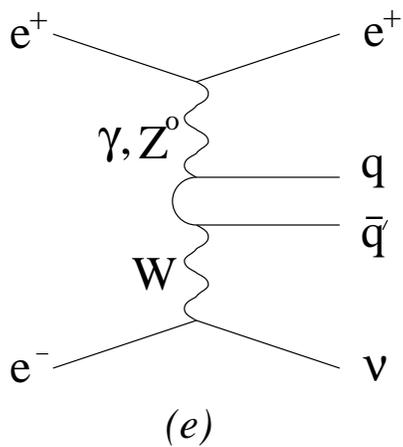
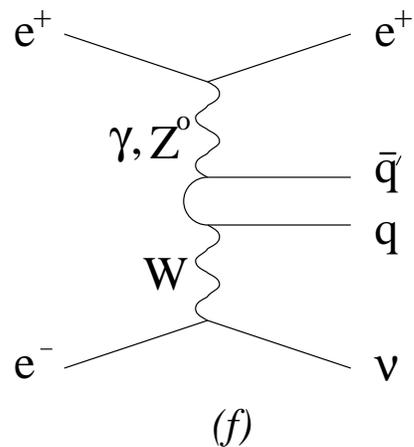
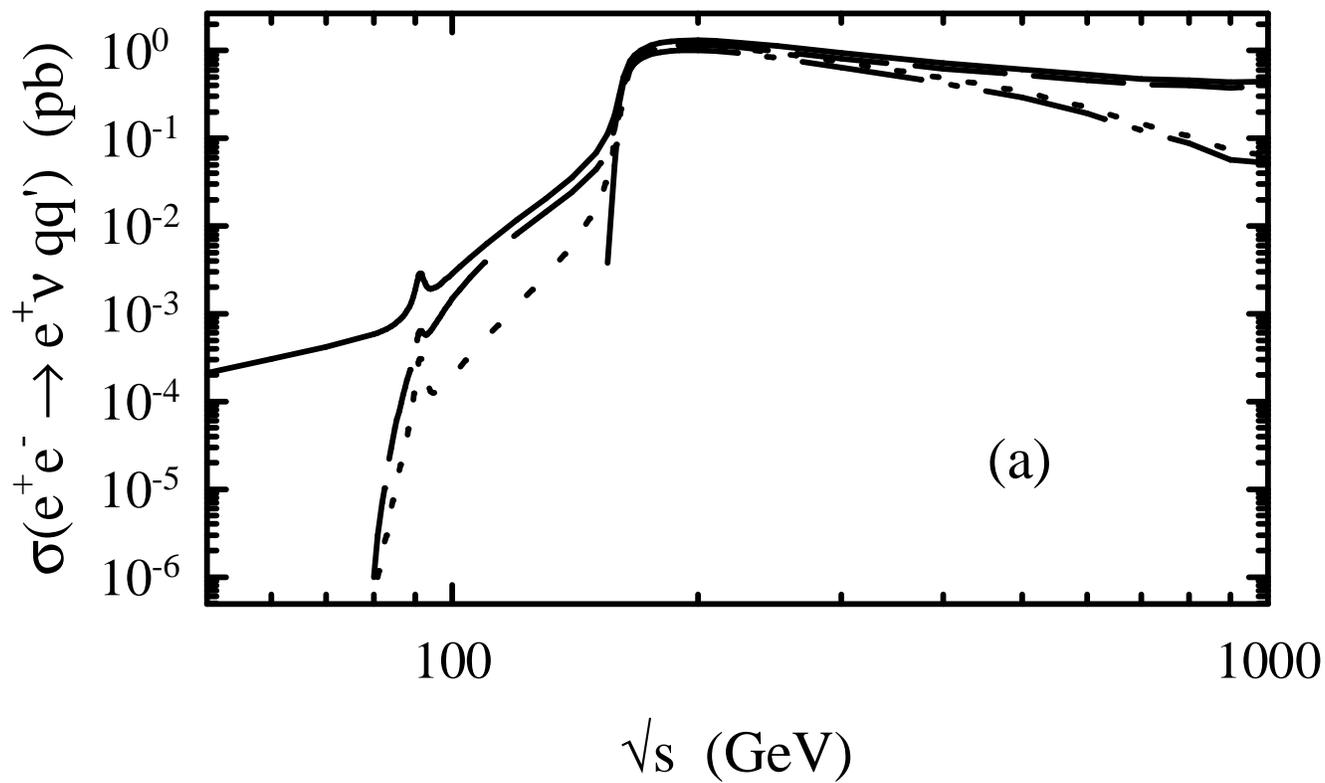
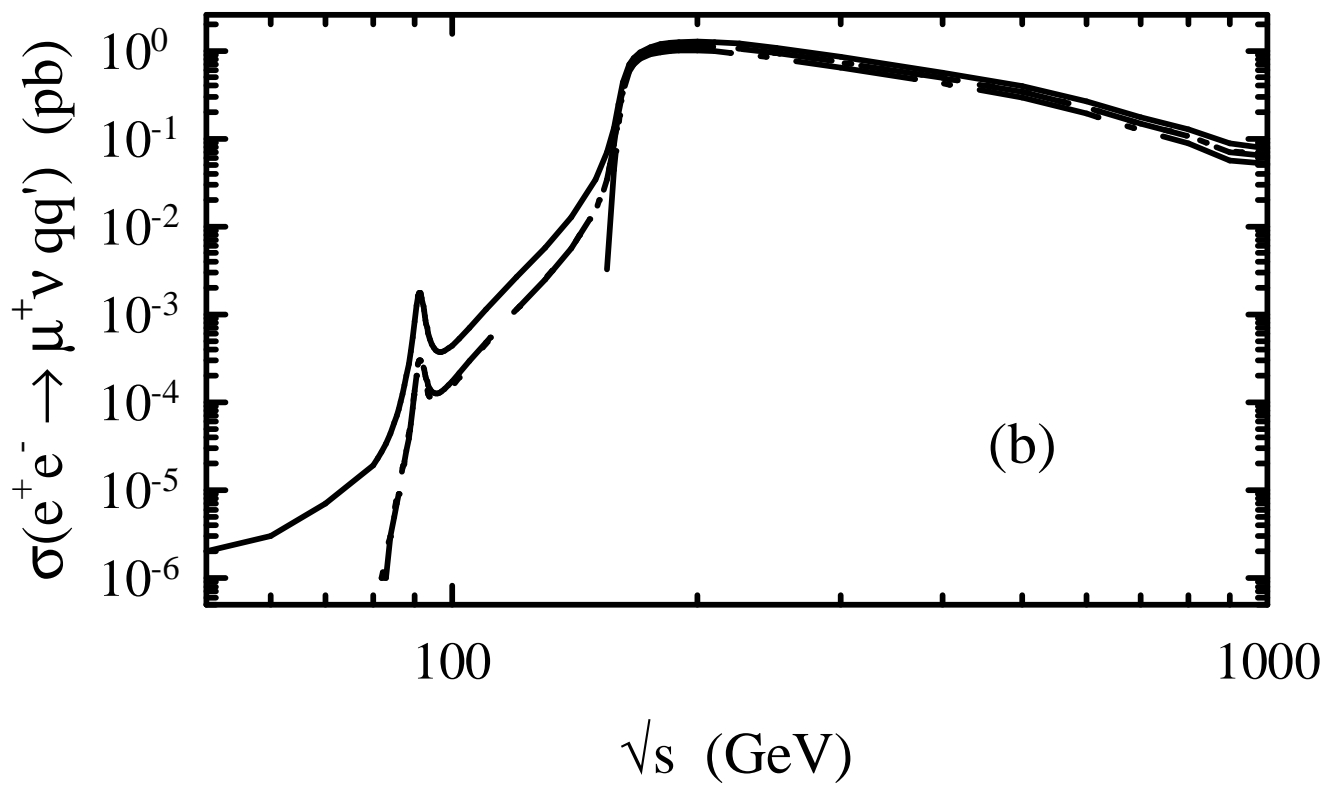
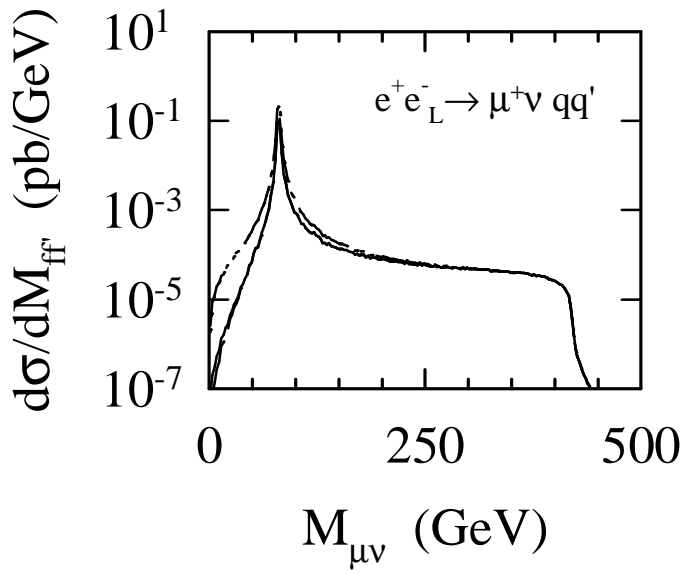
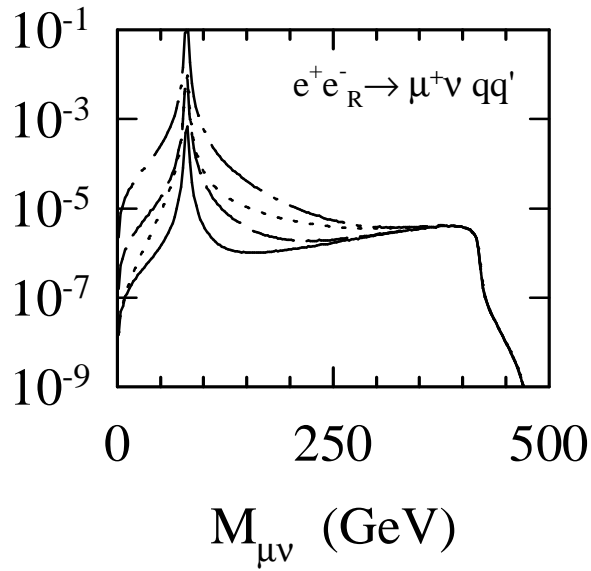
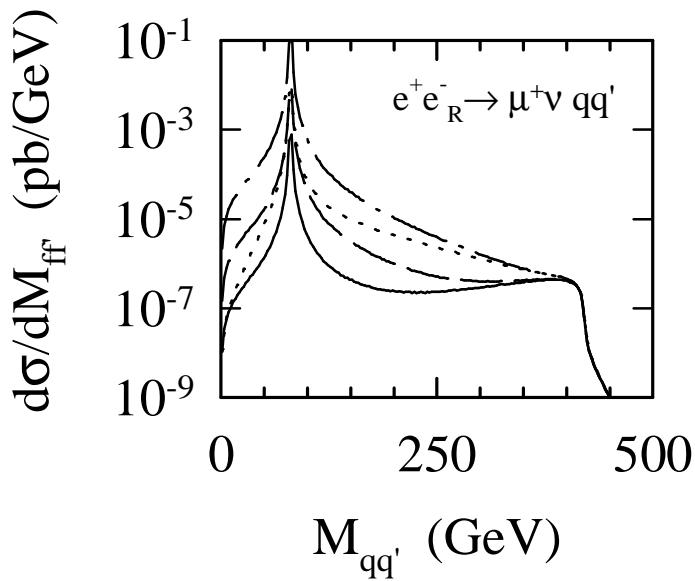
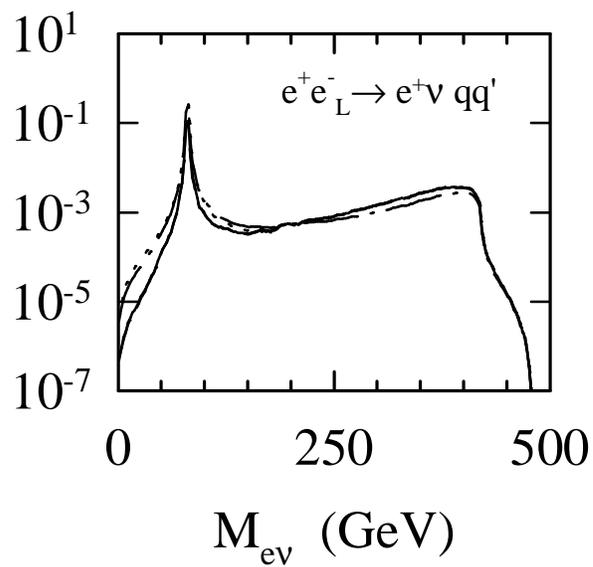
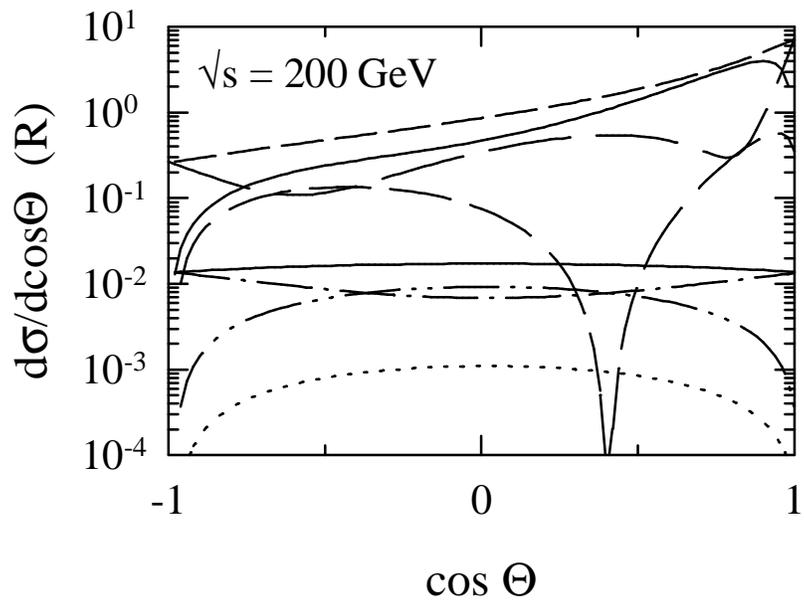
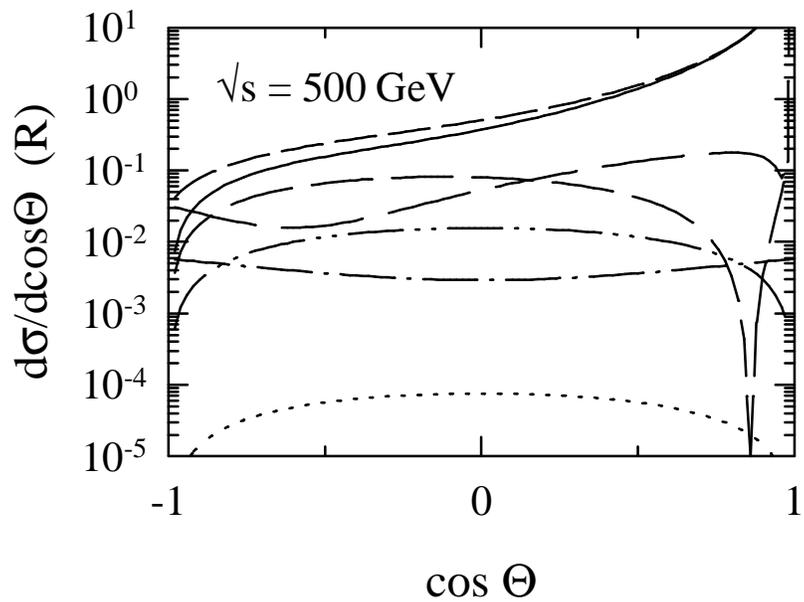
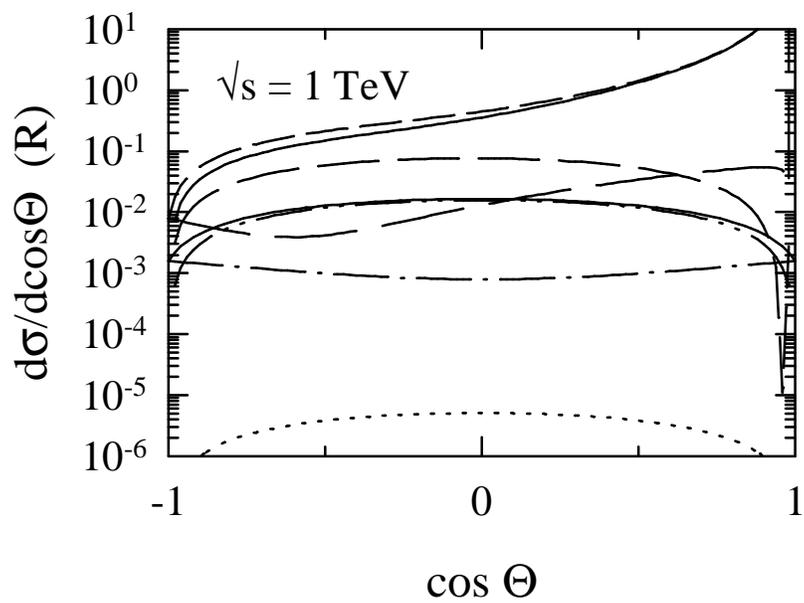
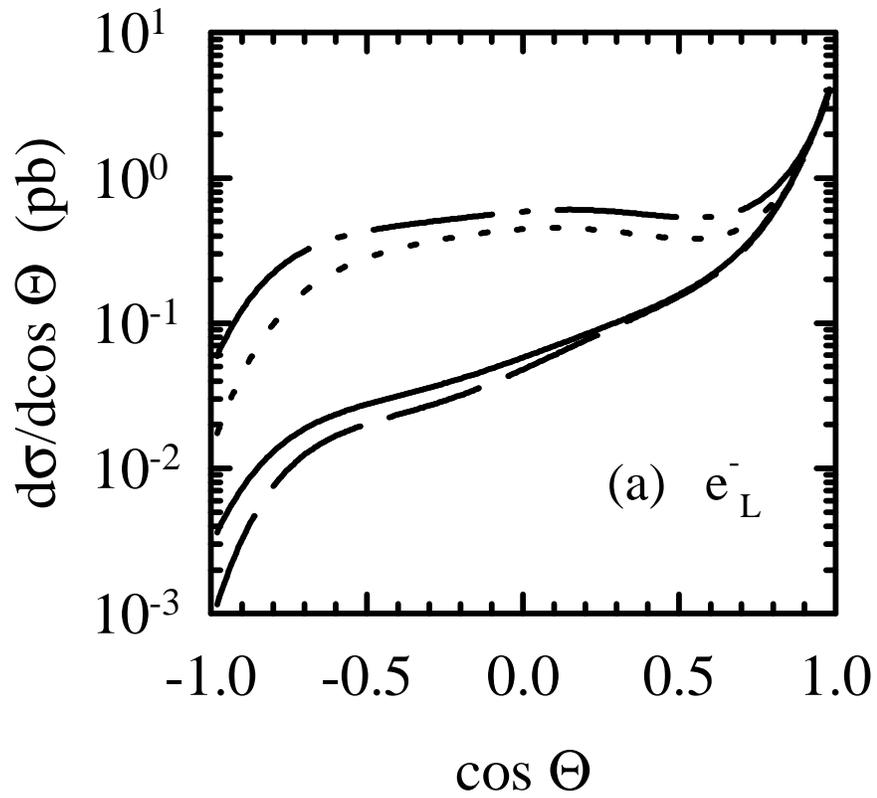
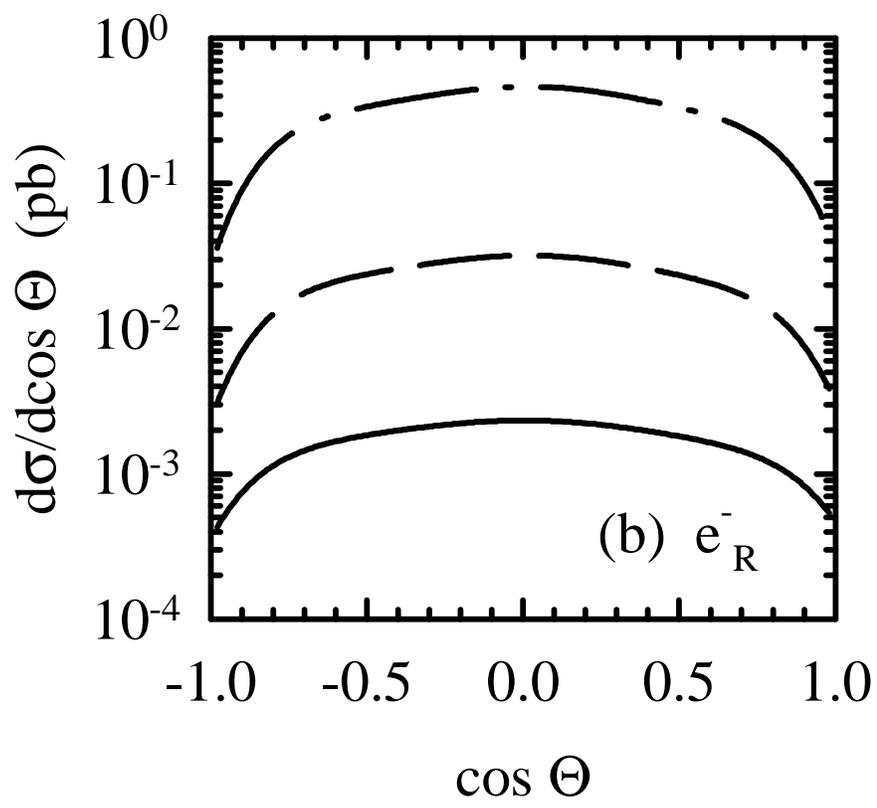
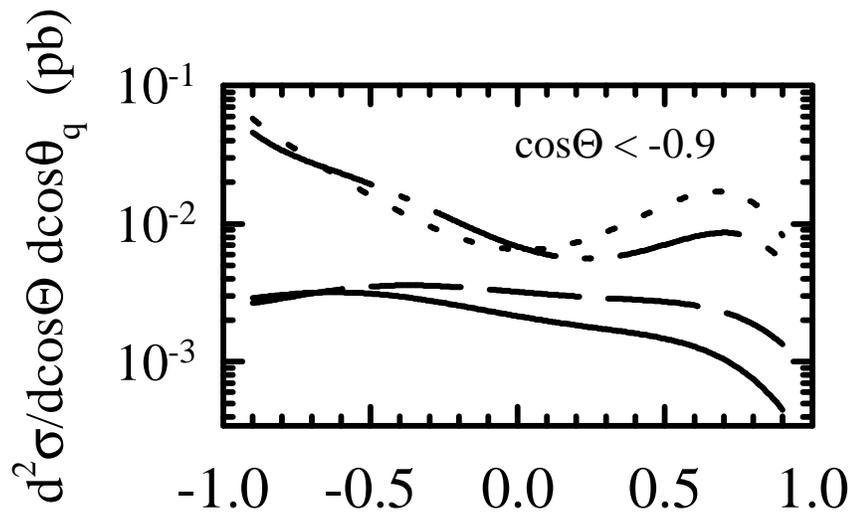
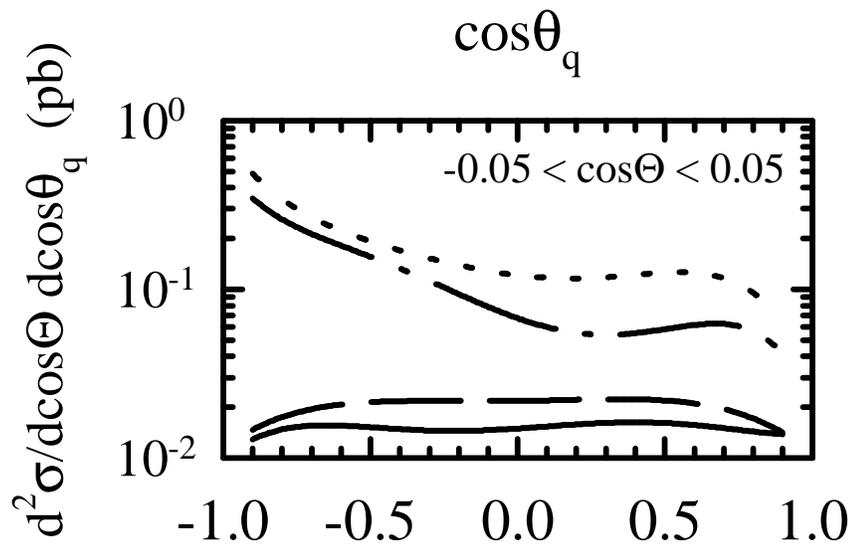
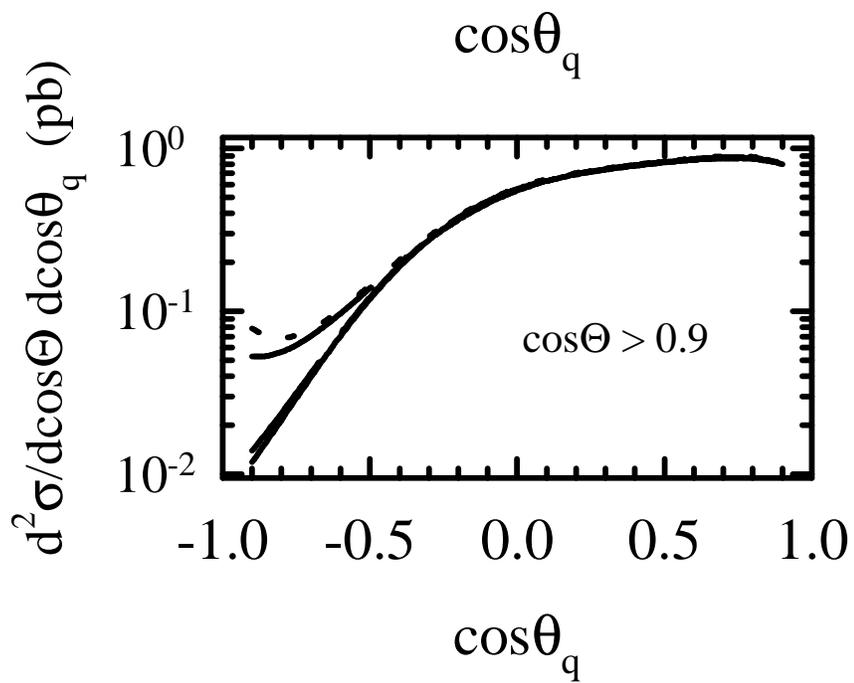
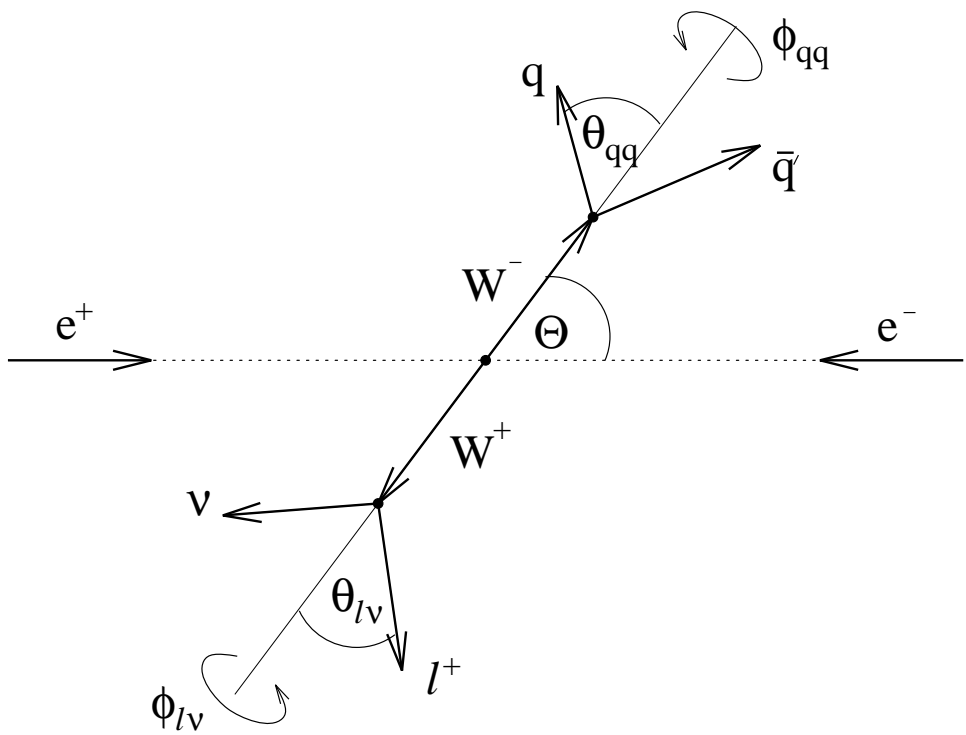
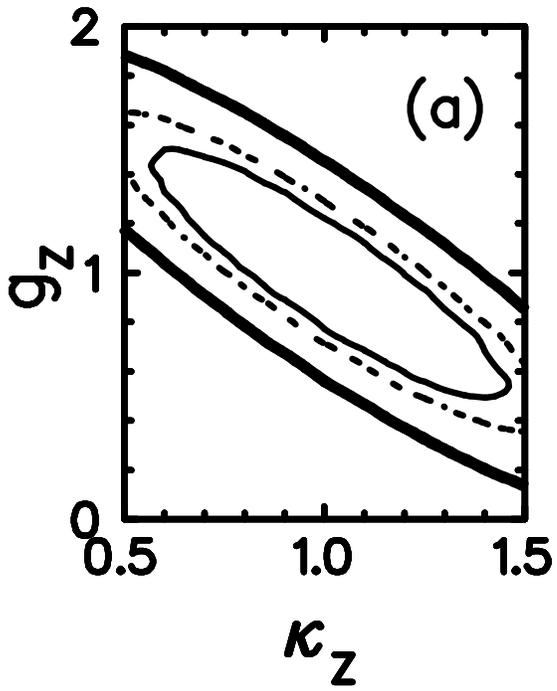
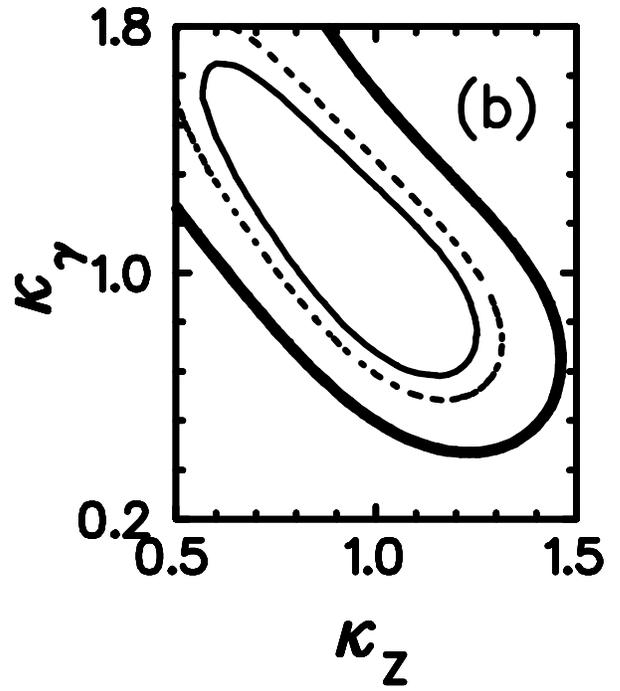
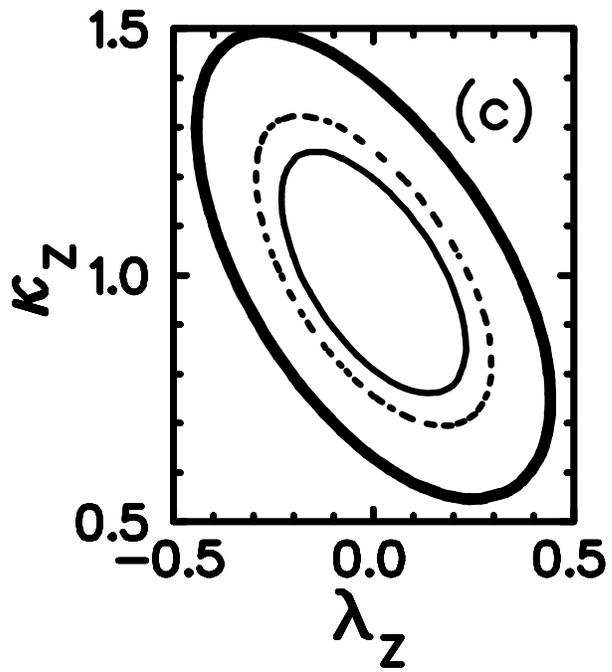
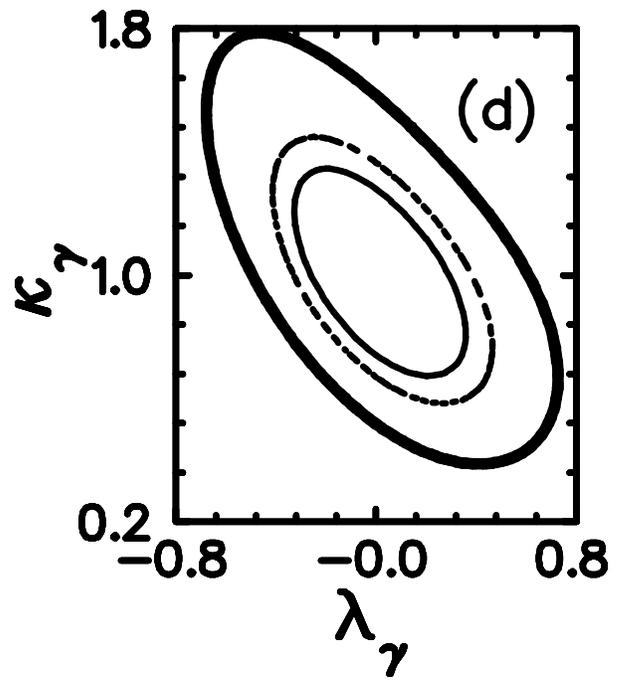
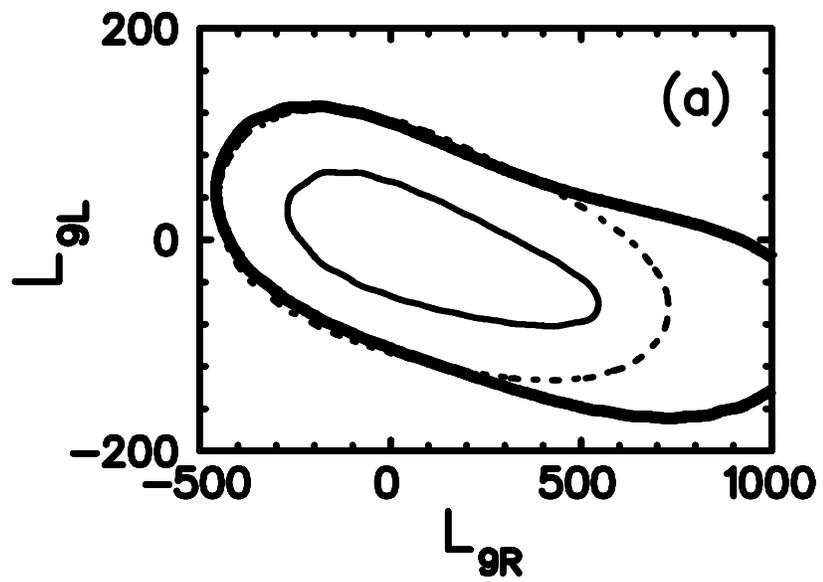
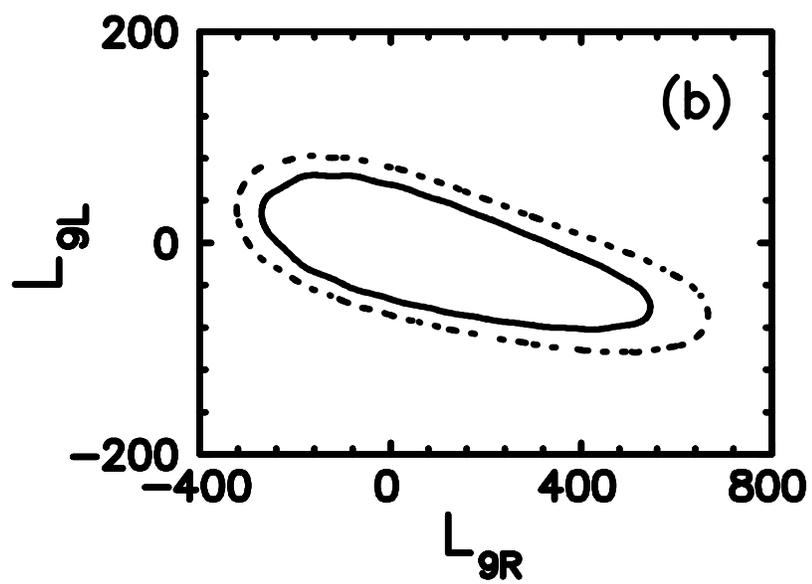
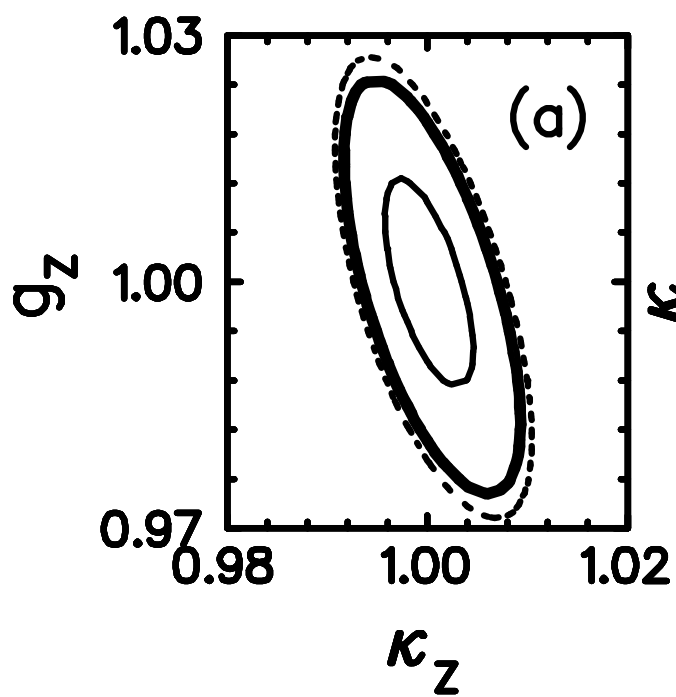
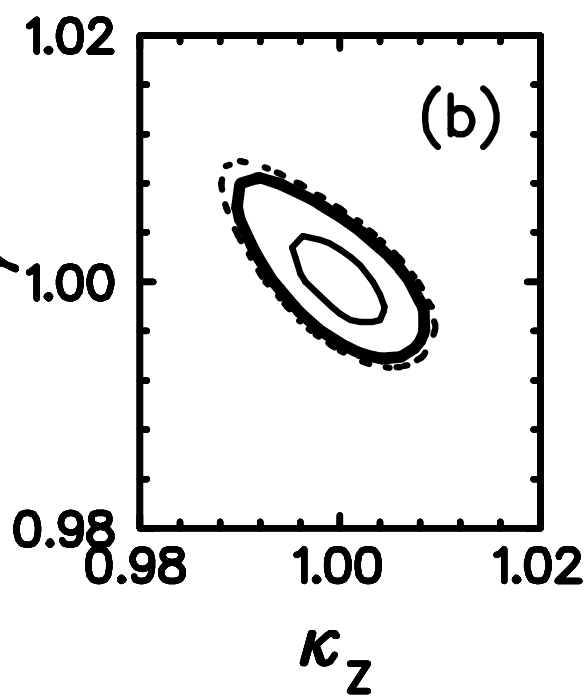
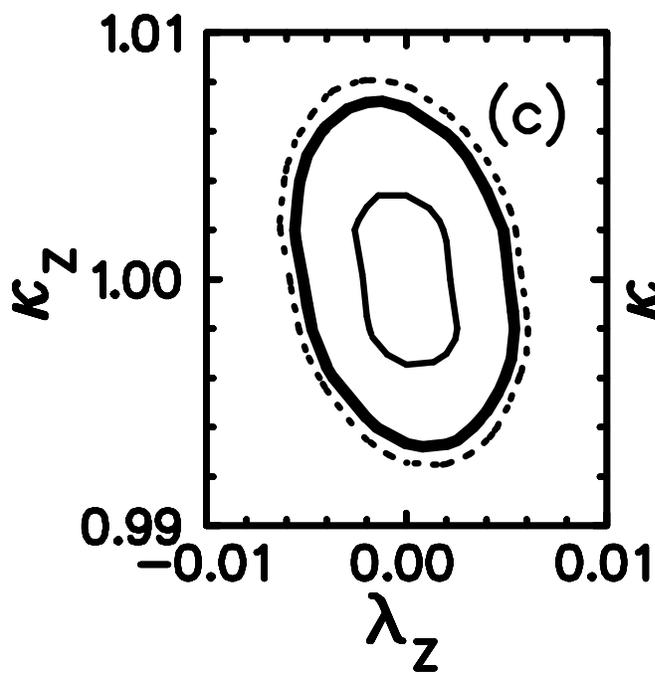
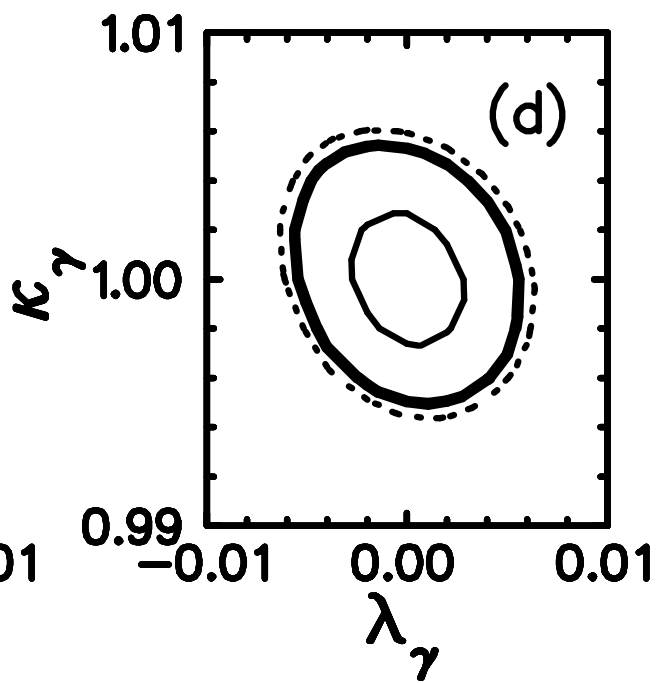
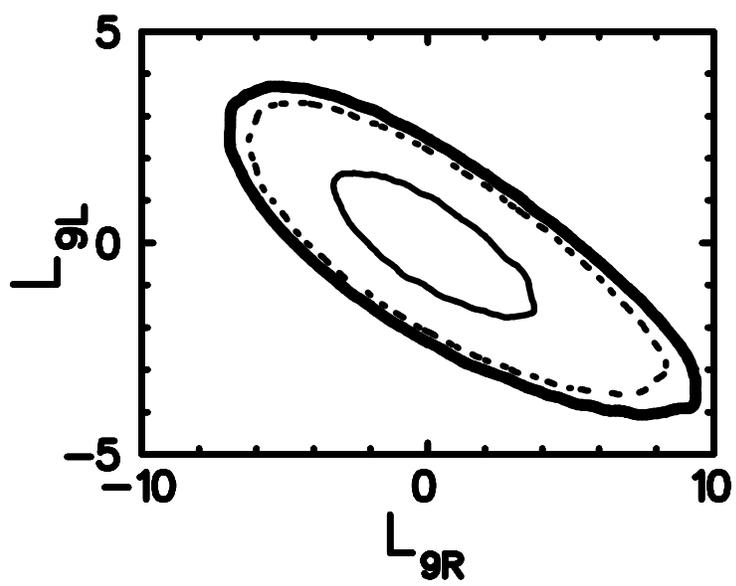
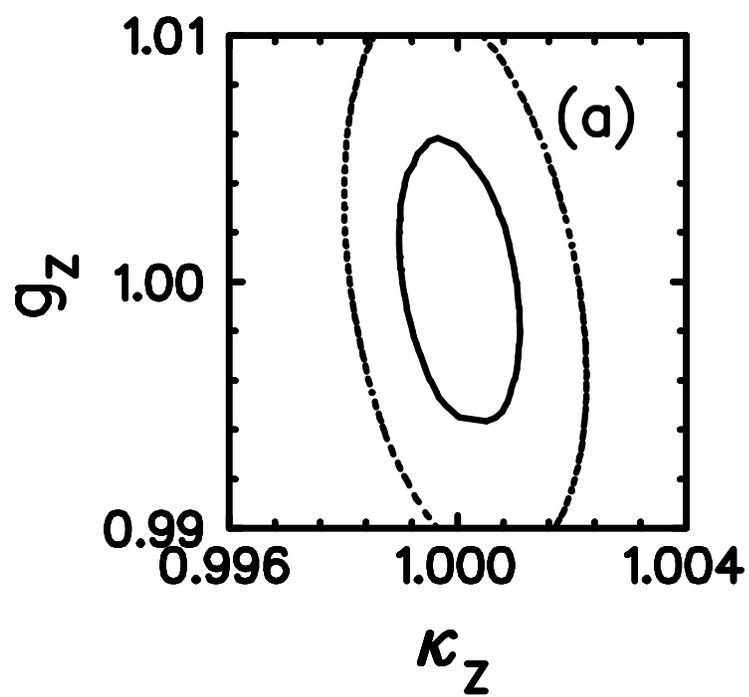
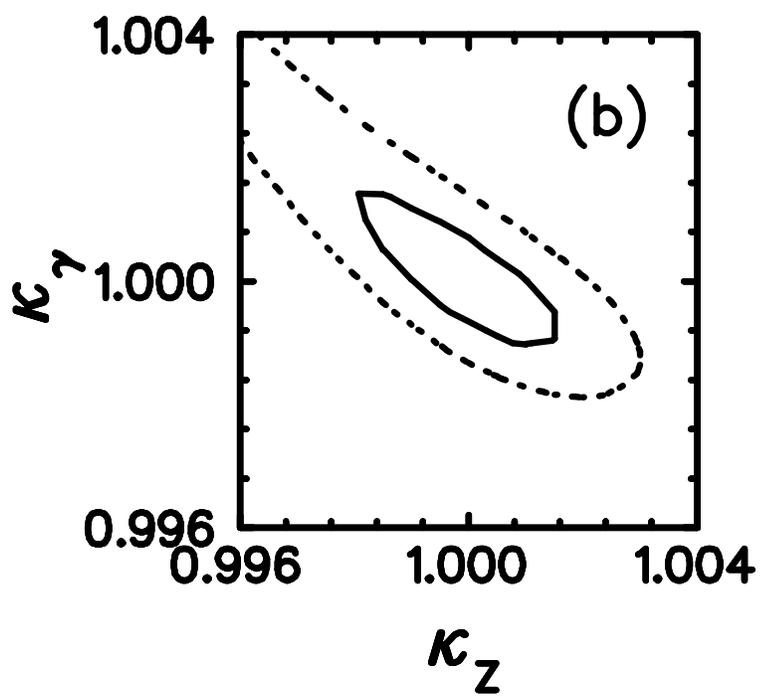
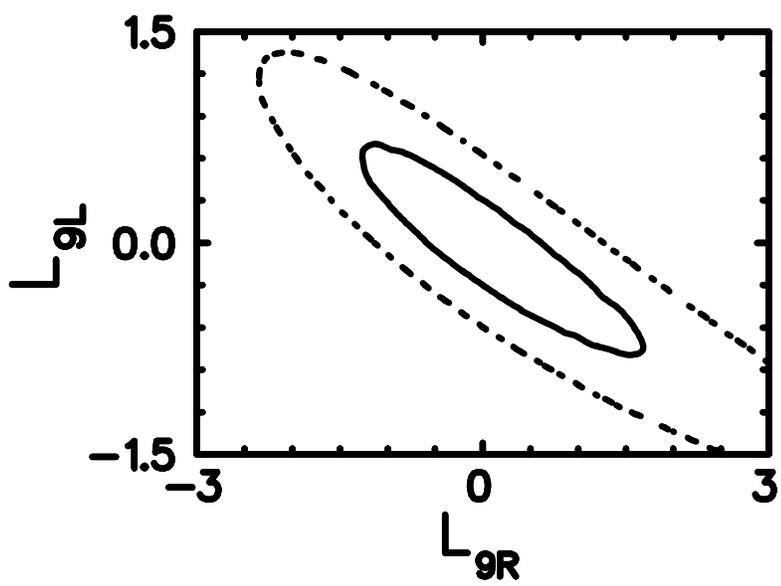
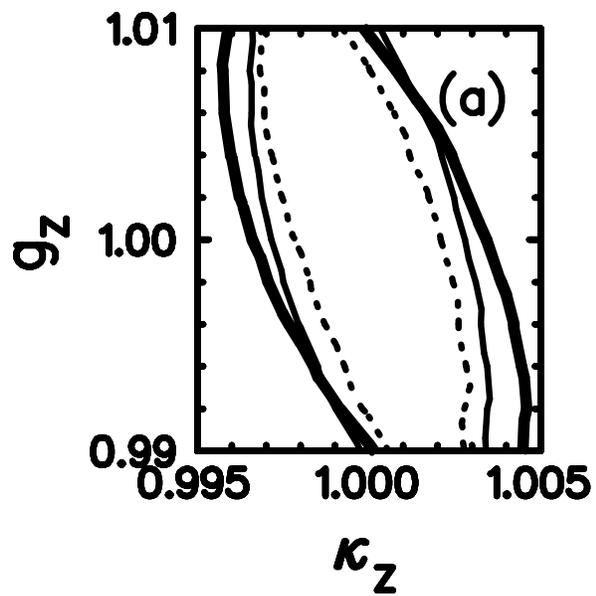
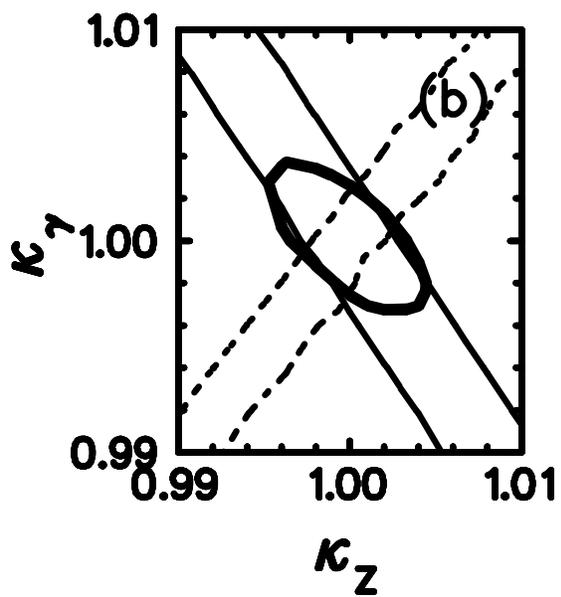
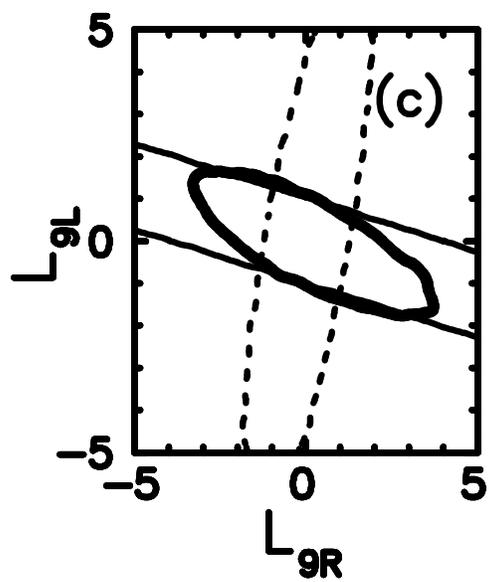
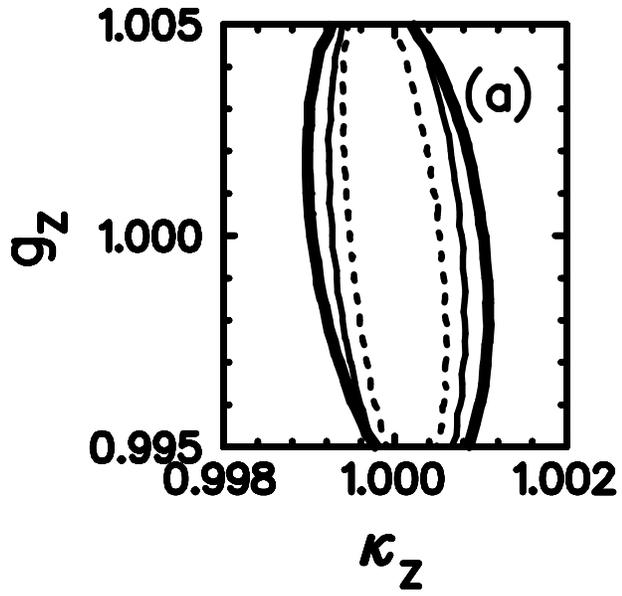
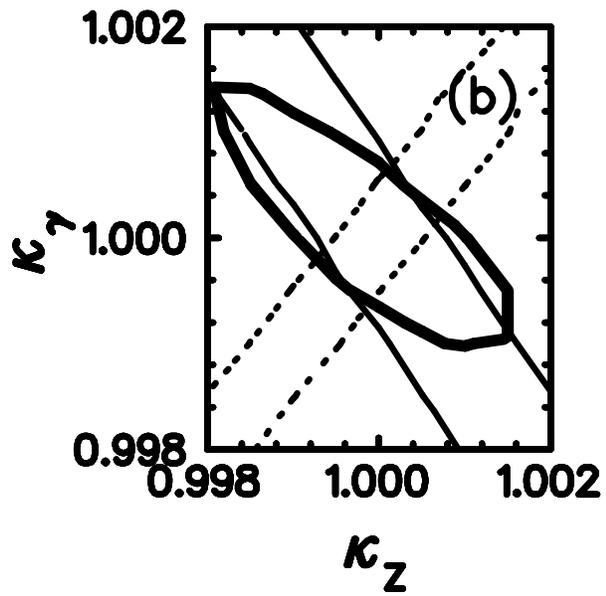
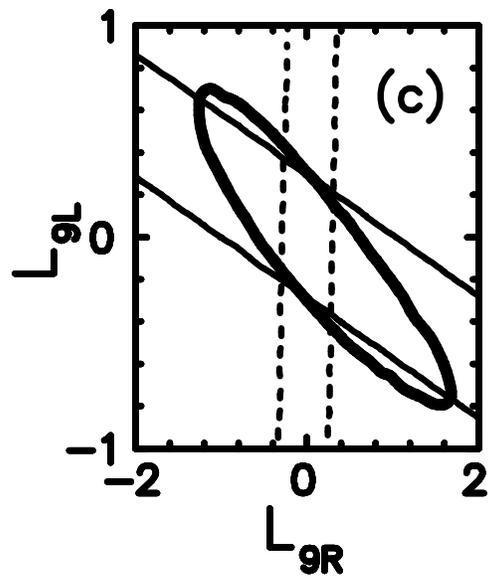
\end{document}